\documentclass[12pt]{iopart}
\usepackage[dvips]{color}
\usepackage{iopams,amsfonts,amssymb,latexsym,epsf,bm} 
\expandafter\let\csname equation*\endcsname\relax
\expandafter\let\csname endequation*\endcsname\relax
\usepackage{amsmath}
\usepackage[dvipdfmx]{graphicx}
\usepackage[hypertex,breaklinks=true,colorlinks=false]{hyperref}
\begin{document}


\title[]{
Excitation band topology and edge matter waves in Bose-Einstein condensates in optical lattices
}
\author{Shunsuke Furukawa$^{1}$ and Masahito Ueda$^{1,2}$}
\address{
$^1$Department of Physics, University of Tokyo, 7-3-1 Hongo, Bunkyo-ku, Tokyo 113-0033, Japan\\
$^2$RIKEN Center for Emergent Matter Science (CEMS), Wako, Saitama 351-0198, Japan
}
\ead{furukawa@cat.phys.s.u-tokyo.ac.jp}

\begin{abstract}
We show that Bose-Einstein condensates in optical lattices with broken time-reversal symmetry 
can support chiral edge modes originating from nontrivial bulk excitation band topology. 
To be specific, we analyze a Bose-Hubbard extension of the Haldane model, 
which can be realized with recently developed techniques of manipulating honeycomb optical lattices. 
The topological properties of Bloch bands known for the noninteracting case 
are smoothly carried over to Bogoliubov excitation bands for the interacting case. 
We show that the parameter ranges that display topological bands enlarge with increasing the Hubbard interaction or the particle density. 
In the presence of sharp boundaries, chiral edge modes appear in the gap between topological excitation bands. 
We demonstrate that by coherently transferring a portion of a condensate into an edge mode, 
a density wave is formed along the edge owing to an interference with the background condensate. 
This offers a unique method of detecting an edge mode through a macroscopic quantum phenomenon. 
\end{abstract}

\pacs{03.75.Kk, 03.65.Vf, 73.43.Cd}


\vspace{2pc}
\noindent{\it Keywords}: Bose-Einstein condensation, Bose-Hubbard model, band topology, chiral edge state
\tableofcontents


\newcommand{\cutspace}{}

\newcommand{\bra}[1]{\langle #1|}
\newcommand{\ket}[1]{|#1 \rangle}
\newcommand{\bracket}[2]{\langle #1|#2 \rangle}
\newcommand{\bbra}[1]{\langle\!\langle #1|}
\newcommand{\kett}[1]{|#1 \rangle\!\rangle}
\newcommand{\bbrackett}[2]{\langle\!\langle #1|#2 \rangle\!\rangle}
\newcommand{\ua}{\uparrow}
\newcommand{\da}{\downarrow}

\newcommand{\Scal}{{\cal S}}
\newcommand{\group}{\mathrm{group}}
\newcommand{\ThD}{\mathrm{(3D)}}

\newcommand{\vout}{v_\mathrm{out}}
\newcommand{\vin}{v_\mathrm{in}}

\newcommand{\rv}{{\bm{r}}}
\newcommand{\kv}{{\bm{k}}}
\newcommand{\Sv}{\bm{S}}
\newcommand{\Rv}{\bm{R}}
\newcommand{\pv}{\bm{p}}
\newcommand{\qv}{\bm{q}}
\newcommand{\Kv}{{\bm{K}}}
\newcommand{\piv}{\bm{\pi}}
\newcommand{\Lv}{\bm{L}}
\newcommand{\ellv}{\bm{\ell}}
\newcommand{\Av}{\bm{A}}
\newcommand{\Bv}{\bm{B}}
\newcommand{\Omegav}{\bm{\Omega}}
\newcommand{\deltav}{{\bm{\delta}}}
\newcommand{\av}{{\bm{a}}}
\newcommand{\bv}{{\bm{b}}}
\newcommand{\hv}{{\bm{h}}}
\newcommand{\sigmav}{{\bm{\sigma}}}
\newcommand{\wv}{{\bm{w}}}
\newcommand{\fv}{{\bm{f}}}
\newcommand{\alphav}{{\bm{\alpha}}}
\newcommand{\betav}{{\bm{\beta}}}

\newcommand{\ah}{\hat{a}}
\newcommand{\bh}{\hat{b}}
\newcommand{\hh}{\hat{h}}
\newcommand{\zh}{\hat{z}}
\newcommand{\xh}{\hat{x}}
\newcommand{\yh}{\hat{y}}
\newcommand{\ph}{\hat{p}}
\newcommand{\ellh}{\hat{\ell}}
\newcommand{\Hh}{\hat{H}}
\newcommand{\Lh}{\hat{L}}
\newcommand{\Nh}{\hat{N}}
\newcommand{\Vh}{\hat{V}}
\newcommand{\Psih}{\hat{\Psi}}
\newcommand{\pdif}[1]{ \frac{\partial}{\partial #1} }
\newcommand{\pdifdif}[1]{ \frac{\partial^2}{\partial #1^2} }
\newcommand{\pdiff}[2]{ \frac{\partial #1}{\partial #2} }

\newcommand{\nt}{\tilde{n}}
\newcommand{\psit}{\tilde{\psi}}
\newcommand{\Psit}{\tilde{\Psi}}
\newcommand{\Vt}{\tilde{V}}
\newcommand{\Zbb}{\mathbb{Z}}
\newcommand{\Rbb}{\mathbb{R}}
\newcommand{\bvec}[2]{\begin{bmatrix} #1 \\ #2\end{bmatrix}}
\newcommand{\Pf}{\mathrm{Pf}}
\newcommand{\Nb}{ {\bar{N}} }
\newcommand{\ex}{\mathrm{ex}}

\newcommand{\gt}{\tilde{g}}
\newcommand{\tot}{\mathrm{tot}}

\newcommand{\ev}{\bm{e}}
\newcommand{\Lambdav}{\bm{\Lambda}}
\newcommand{\Nuc}{{N_\mathrm{uc}}}
\newcommand{\nuc}{{n_\mathrm{uc}}}


\section{Introduction}\label{sec:intro}


Topological insulators and superconductors have attracted great attention in recent years 
for their rich variety of quantized responses and robust gapless edge states 
originating from nontrivial topology of bulk Bloch bands \cite{Hasan10,Qi11,Bernevig13}. 
A prototype of topological insulators is an integer quantum Hall system \cite{vonKlitzing86}, 
which exhibits a quantized Hall conductivity $\sigma_{xy}=-(e^2/h) C$, 
where $C$ is the sum of the Chern numbers of occupied bands \cite{Thouless82}. 
The gapless edge states are characterized by $|C|$ sets of chiral modes propagating clockwise (counterclockwise) along the system's edge for $C>0$ ($C<0$) \cite{Halperin82,Hatsugai93}. 
Here a nonzero value of $C$ is caused by the breaking of time-reversal symmetry due to a magnetic field. 
The discovery of  $\mathbb{Z}_2$ topological insulators \cite{Kane05,Bernevig06,Konig07,Fu07,Moore07,Roy09,Hsieh08} 
has opened up a new avenue for realizing a topologically nontrivial structure in Bloch bands through spin-orbit coupling, without breaking time-reversal symmetry. 
Topological superconductors have been shown to exhibit exotic edge states consisting of Majorana fermions, which are protected by particle-hole symmetry \cite{Read00,Ivanov01}. 
A unified understanding of these topological phases has been achieved with a topological periodic table, 
where such phases are systematically classified for quadratic fermionic Hamiltonians in different dimensions and symmetry classes \cite{Schnyder08,Kitaev09}. 


Ultracold atomic systems have recently emerged as a new platform for exploring the physics of topological phases,  
especially owing to ongoing experimental developments for engineering synthetic gauge fields \cite{Dalibard11,Goldman13} which can be used to produce such states. 
Different schemes have been proposed and implemented to create nearly uniform magnetic fields in continuum \cite{Lin09}, 
optical lattices \cite{Aidelsburger13,Miyake13,Aidelsburger15,Kennedy15}, 
and synthetic dimensions \cite{Celi14,Mancini15,Stuhl15}. 
Furthermore, the Haldane model \cite{Haldane88}, in which non-uniform fluxes pierce through the system, 
has been realized using fermionic atoms in a periodically modulated honeycomb optical lattice \cite{Jotzu14} 
(see Refs.~\cite{Shao08,Stanescu09,Alba11,Goldman13NJ,Zheng14} for early theoretical proposals for realizing the same or related models). 
The Haldane model is a prototypical example of Hamiltonians 
that exhibit topologically distinct regimes characterized by the Chern numbers $C_\pm$ associated with upper ($+$) and lower ($-$) Bloch bands. 
The phase diagram of this model 
has been vindicated experimentally using momentum-resolved interband transitions \cite{Jotzu14}. 
Another recent remarkable achievement has been an interferometric measurement of the $\pi$ Berry flux in the momentum space of a honeycomb lattice \cite{Duca15}. 


A notable feature of atomic systems is that one can study the effect of quantum statistics. 
For example, by implementing the technique of Ref.~\cite{Jotzu14} for bosonic atoms, 
one can realize a bosonic counterpart of the Haldane model. 
In the noninteracting case, topological properties of Bloch bands do not depend on quantum statistics. 
For weakly interacting bosons in optical lattices, 
Bogoliubov excitation bands give the elementary excitations of Bose-Einstein condensates (BEC). 
It is then interesting to ask how the topological properties of Bloch bands are carried over to those of Bogoliubov excitation bands in the interacting case. 


Band topology of bosonic or classical vibrational modes has been studied previously 
in photonic \cite{Haldane08,ZhangNiu10,Wang09,Hafezi13,Rechtsman13}, phononic \cite{Prodan09,Berg11,Susstrunk15}, 
magnonic \cite{Shindou13,Shindou14,Romhanyi15}, and polaritonic \cite{Karzig15} excitations. 
Nontrivial Chern numbers of bulk excitation bands give rise to 
in-gap chiral edge modes, as observed experimentally in photonic systems \cite{Wang09,Hafezi13,Rechtsman13}. 
Ultracold bosonic atoms in optical lattices are expected to offer a unique platform for the studies of band topology 
because of the high controllability of such systems and a potential combination with the macroscopic quantum nature of BECs. 


In this paper, we study the topological properties of Bogoliubov excitation bands in BECs in optical lattices with broken time-reversal symmetry,  
using a Bose-Hubbard extension of the Haldane model (Haldane-Bose-Hubbard model). 
We show that the topological properties of the Bloch bands in the noninteracting case \cite{Haldane88}
are smoothly carried over to those of the Bogoliubov excitation bands in the interacting case. 
Furthermore, the parameter ranges that exhibit nontrivial band topology enlarge with increasing the Hubbard interaction or the particle density (see Fig.~\ref{fig:phase}). 
In the presence of sharp boundaries, chiral edge modes appear in the gap between topologically nontrivial excitation bands. 
We demonstrate that by coherently transferring a portion of the condensate into an edge mode, 
a density wave is formed along the edge owing to an interference with the background condensate. 
This property can be used as a macroscopically enhanced signature of an edge mode. 

We note that Vasic {\it et al.}\ \cite{Vasic15} have recently studied the ground-state phase diagram of the Haldane-Bose-Hubbard model, 
predicting the emergence of uniform and chiral BEC phases and plaquette Mott insulators with loop currents. 
While the Bogoliubov excitation bands in the uniform and chiral BEC phases have also been studied, 
their topological properties such as Chern numbers and associated edge modes have not been analyzed in detail. 
Our work addresses such topological properties of excitations in a uniform BEC phase, 
clarifies the parameter ranges showing topologically nontrivial bands in the presence of interactions, 
and demonstrates a unique method of detecting an edge mode though a macroscopic quantum interference. 
We also note that strong correlation effects on the band topology have been studied 
in the spin-$\frac12$ fermionic Haldane-Hubbard model in Refs.~\cite{HeJing11,Prychynenko15,ZhengWei15}. 


The rest of the paper is organized as follows. 
In Sec.~\ref{sec:model}, we describe our model, and review the band structure in the noninteracting case. 
In Sec.~\ref{sec:Bogoliubov}, we present a Bogoliubov theory for homogeneous condensates with weak repulsive interactions, 
and analyze the topology of Bogoliubov excitation bands. 
In Secs.~\ref{sec:inhomo_BEC_formalism} and \ref{sec:inhomo_BEC_numerics}, 
we analyze the ground state and excitations of inhomogeneous condensates 
by using a Bogoliubov-de Gennes theory. 
After describing the basic formalism in Sec.~\ref{sec:inhomo_BEC_formalism}, 
we present numerical results for box and harmonic traps in Sec.~\ref{sec:inhomo_BEC_numerics}. 
In Sec.~\ref{sec:summary}, we present a summary of this paper and discuss an outlook for future studies. 

\section{Model and band structure} \label{sec:model}

In this section, we first describe our model---a Bose-Hubbard version of the Haldane model in a honeycomb lattice. 
We then review the band structure in the noninteracting case, 
and discuss the single-particle ground state into which bosons condense at zero temperature. 

\subsection{Bose-Hubbard Hamiltonian}\label{sec:BH}


We consider bosonic atoms in an optical lattice, 
which are well described in the tight-binding limit by a Bose-Hubbard model. 
The Hamiltonian of our system is given by 
\begin{equation}\label{eq:H}
 H = -\sum_{\rv,\rv'} J(\rv,\rv') a^\dagger(\rv) a(\rv') + \frac{U}2 \sum_\rv a^\dagger(\rv)^2 a(\rv)^2, 
\end{equation}
where $\rv$ and $\rv'$ run over all the site positions of the lattice,  
$a(\rv)$ is the bosonic annihilation operator at the site $\rv$, and $U$ describes the on-site Hubbard interaction.  
The diagonal element $J(\rv,\rv)$ gives a potential energy at the site $\rv$,  
and the off-diagonal element $J(\rv,\rv')$ with $\rv\ne\rv'$ describes the (generally complex) hopping amplitude between the two sites  
and satisfies $J(\rv',\rv)=J^*(\rv,\rv')$ in order for $H$ to be hermitian. 
We set the total number of particles to $N$:
\begin{equation}\label{eq:particle_N}
 \sum_\rv a^\dagger(\rv) a(\rv)=N. 
\end{equation}

\begin{figure}
\begin{center}\includegraphics[width=0.7\textwidth]{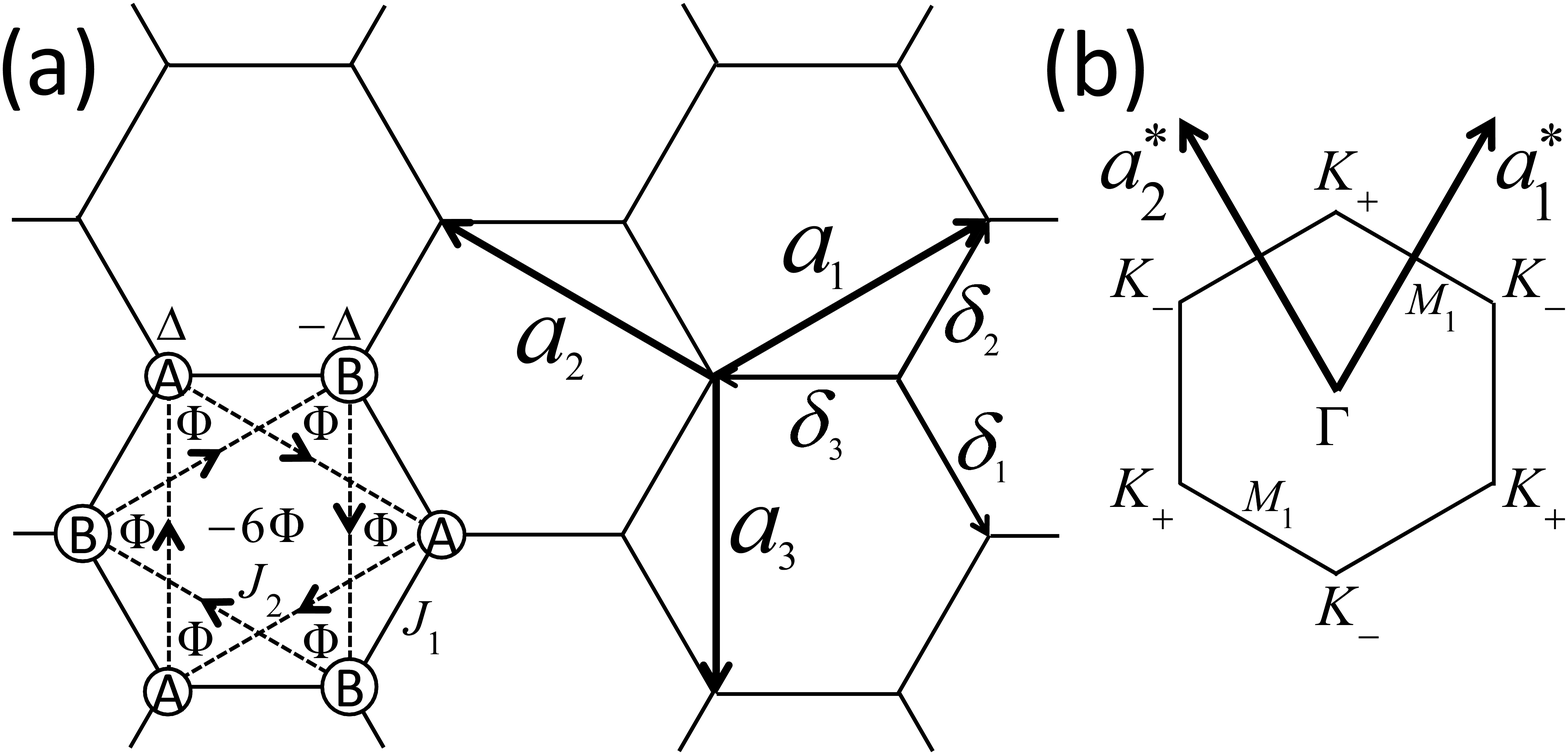}
\cutspace
\caption{\label{fig:model}
(a) Haldane model on a honeycomb lattice.
The kinetic part of the model, $H_\mathrm{kin}$, 
consists of the nearest-neighbor hopping $J_1$, the next-nearest-neighbor hopping $J_2$, 
and the potential difference $2\Delta$ between the two sublattices $A$ and $B$. 
Furthermore, atoms acquire an Aharanov-Bohm phase $\Phi$ 
when hopping along every dashed line in the arrowed direction [see Eq.~\eqref{eq:J_Haldane}]. 
(b) The first Brillouin zone of the honeycomb lattice. 
Here $\av_1^*$ and $\av_2^*$ represent the reciprocal lattice vectors [see Eq.~\eqref{eq:aj_s}].
}\end{center}
\end{figure}


We focus on the case in which the hopping terms in Eq.~\eqref{eq:H}, which will be denoted by $H_\mathrm{kin}$, are given 
by the Haldane model in a honeycomb lattice \cite{Haldane88}; see Fig.~\ref{fig:model}. 
A honeycomb lattice consists of $A$ and $B$ sublattices, and is spanned by the primitive vectors $\av_{1,2}$ 
($\av_3$ is also introduced for convenience). 
We also introduce the vectors $\deltav_{1,2,3}$, which are directed from a $B$ site to the three neighboring $A$ sites. 
These vectors are given by
\begin{equation}\label{eq:delj_aj}
\begin{split}
 &\deltav_1=d \left(\frac12,-\frac{\sqrt{3}}{2}\right),~
 \deltav_2=d \left(\frac12,+\frac{\sqrt{3}}{2}\right),~
 \deltav_3=d(-1,0),\\
 &\av_1=\deltav_2-\deltav_3,~\av_2=\deltav_3-\deltav_1,~\av_3=\deltav_1-\deltav_2, 
\end{split}
\end{equation}
where $d$ is the length between the neighboring sites. 
We also introduce the reciprocal lattice vectors
\begin{equation}\label{eq:aj_s}
 \av_1^* = \frac{2\pi}{d} \left(  \frac13,\frac1{\sqrt{3}} \right), ~
 \av_2^* = \frac{2\pi}{d} \left( - \frac13,\frac1{\sqrt{3}} \right), 
\end{equation}
which satisfy $\av_i\cdot\av_j^*=2\pi \delta_{ij}~(i,j=1,2)$. 

The Haldane model consists of the real nearest-neighbor hopping $J_1$, the complex next-nearest-neighbor hopping $J_2e^{\pm i\Phi}$, 
and the potential difference $2\Delta$ between the two sublattices. 
Nonzero values of $J(\rv,\rv')$ are thus given by
\begin{equation}\label{eq:J_Haldane}
\begin{split}
 &-J(\rv+\deltav_j,\rv)=-J(\rv,\rv+\deltav_j)=-J_1~~\text{for}~\rv\in B; \\
 &-J(\rv+\av_j,\rv)=-J^*(\rv,\rv+\av_j)=-J_2 e^{-i\epsilon_X \Phi}~~\text{for}~\rv\in X=A,B; \\
 &-J(\rv,\rv)=\epsilon_X \Delta+V(\rv)~~\text{for}~\rv\in X=A,B,
\end{split}
\end{equation}
where $j=1,2,3$, and $\epsilon_{A,B}=\pm1$. 
Here, $\rv\in X$ indicates that the site $\rv$ belongs to the $X$ sublattice,  
and $V(\rv)$ describes an external potential which depends on the setting of our system. 
We assume $J_1,J_2>0$ in the following. 

\subsection{Band structure in the noninteracting case}\label{sec:band}

\begin{figure}
\begin{center}\includegraphics[width=0.8\textwidth]{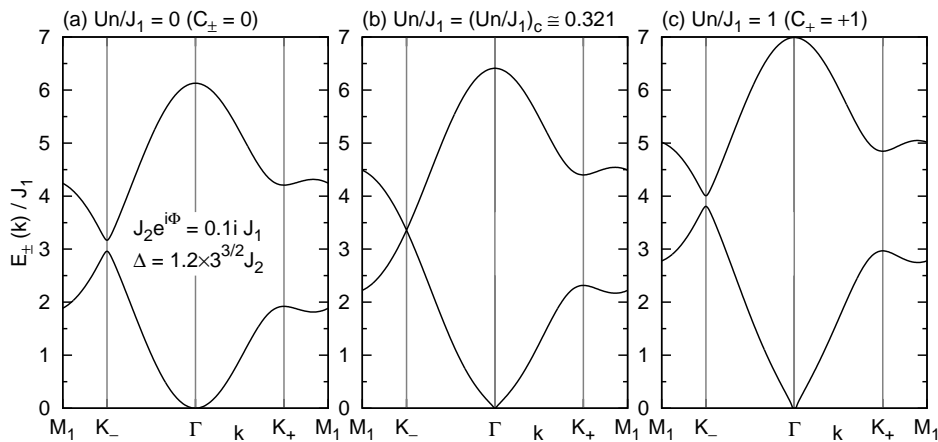}
\cutspace
\caption{(color online) \label{fig:bands}
(a) An example of a band structure in the noninteracting case, 
calculated along the path $M_1\to K_-\to K_+ \to M_1$ in the first Brillouin zone shown in Fig.~\ref{fig:model}(b). 
The parameters are chosen to be in the trivial phase with $C_\pm=0$. 
(b, c) Examples of Bogoliubov excitation bands in the presence of interaction. 
A transition in the band topology occurs at $Un/J_1=(Un/J_1)_\mathrm{c}\simeq0.321$ [a solution to Eq.~\eqref{eq:phase_HBH_exact}], 
at which the band gap closes at the $K_-$ point [see (b)]. 
For $Un/J_1>(Un/J_1)_\mathrm{c}$, the higher band acquires a nontrivial Chern number $C_+=+1$. 
}\end{center}
\end{figure}

Here we set $U=0$, and review the band structure of the Haldane model in the noninteracting case \cite{Haldane88}. 
Assuming the periodic boundary conditions in the two directions of the honeycomb lattice, 
we perform the Fourier expansion 
\begin{equation}\label{eq:ar_ak}
 a(\rv)=\frac1{\sqrt{\Nuc}} \sum_\kv a_X(\kv) e^{i\kv\cdot\rv}~~(\rv\in X=A,B),
\end{equation}
where the sum is taken over the discrete momenta $\kv$ in the first Brillouin zone, 
and $\Nuc$ is the total number of unit cells in the system (i.e., half of the total number of sites). 
The kinetic part of the Hamiltonian \eqref{eq:H} is then rewritten as
\begin{equation}
 H_\mathrm{kin}=\sum_\kv \left(a_A^\dagger (\kv),a_B^\dagger (\kv) \right) {\cal H}(\kv) 
\begin{pmatrix}
 a_A (\kv) \\ a_B (\kv)
\end{pmatrix}.
\end{equation}
Here the $2\times 2$ hermitian matrix ${\cal H}(\kv)$ can be written in the form
\begin{equation}\label{eq:H_k}
 {\cal H}(\kv)= h_0(\kv) I + \hv(\kv)\cdot\sigmav, 
\end{equation}
where $I$ is the identity matrix, $\sigmav=(\sigma_1,\sigma_2,\sigma_3)$ are the Pauli matrices. 
The coefficients $h_0(\kv)$ and $\hv(\kv)=(h_1(\kv),h_2(\kv),h_3(\kv))$ are calculated as
\begin{equation}\label{eq:h0123}
\begin{split}
 &h_0(\kv) = -2J_2 \cos(\Phi) \sum_j \cos(\kv\cdot\av_j),~~
 h_1(\kv)= -J_1 \sum_j \cos(\kv\cdot\delta_j),\\
 &h_2(\kv)= -J_1 \sum_j \sin(\kv\cdot\delta_j),~~
 h_3(\kv)= \Delta+2J_2\sin (\Phi) \sum_j \sin(\kv\cdot\av_j). 
\end{split}
\end{equation}
The two energy bands are obtained through the diagonalization of Eq.~\eqref{eq:H_k} as
\begin{equation}\label{eq:bands}
 e_\pm (\kv) = h_0(\kv)\pm h(\kv),~~ h(\kv):=|\hv(\kv)|. 
\end{equation}
An example of energy bands is presented in Fig.~\ref{fig:bands}(a). 
For noninteracting fermions, complete filling of the lower band leads to a band insulator. 


For noninteracting bosons, Bose-Einstein condensation into the lowest-energy single-particle state occurs at zero temperature.\footnote{
In the thermodynamic limit, 
Bose-Einstein condensation does not occur at finite temperatures in two dimensions. 
In finite-size systems, however, a large condensate fraction can still be achieved  
if coherence is formed over the system at sufficiently low temperatures. 
} 
When $J_2=\Delta=0$, the bottom of the lower band  $e_-(\kv)$ is located at $\kv=\bm{0}$. 
To find whether the position of the bottom can change owing to finite $J_2$ or $\Delta$, 
we expand Eq.~\eqref{eq:bands} around $\kv=\bm{0}$ as
\begin{equation}
 e_\pm (\kv)
 =h_0(\bm{0}) + \frac92 J_2 k^2 d^2 \cos\Phi 
 \pm\left(h(\bm{0})-\frac94 \frac{J_1^2 k^2 d^2}{\sqrt{9J_1^2+\Delta^2}}\right) +{\cal O}(k^3).
\end{equation}
Therefore, $e_-(\kv)$ is minimized at $\kv=\bm{0}$ if the following condition is met: 
\begin{equation}\label{eq:em_cond}
 \frac{J_1^2}{\sqrt{9J_1^2+\Delta^2}} + 2J_2\cos\Phi >0. 
\end{equation}
This condition is satisfied in the regime $J_2,|\Delta| \ll J_1$, which is relevant to the Haldane model realized 
in the scheme of Ref.~\cite{Jotzu14}.\footnote{
When $\Delta=0$ and $\Phi=\pi$, the condition in Eq.~\eqref{eq:em_cond} is written simply as $J_1>6J_2$.  
By relating phases of bosons to angles of classical XY spins, 
we find that this condition is equivalent to the known stability condition of the ferromagnetic order 
in a honeycomb lattice magnet with competing ferromagnetic $2J_1$ and antiferromagnetic $-2J_2$ couplings \cite{Fouet01,Okuma10}. }

To determine the single-particle ground state, we parametrize $\hv(\bm{0})$ using the polar coordinates as 
\begin{equation}
 \hv (\bm{0}) =(-3J_1,0,\Delta)
 = h(\bm{0}) (\sin\theta_0\cos\pi, \sin\theta_0\sin\pi,\cos\theta_0).
\end{equation}
The $\kv=\bm{0}$ part of $H_\mathrm{kin}$ is then diagonalized by means of the transformation 
\begin{equation}\label{eq:aAB_a0pm}
\begin{pmatrix}
 a_A(\bm{0}) \\ a_B(\bm{0})
\end{pmatrix}
=
U(\theta_0,\pi)
\begin{pmatrix}
 a_+(\bm{0}) \\ a_-(\bm{0})
\end{pmatrix},
\end{equation}
where we have defined the $2\times 2$ unitary matrix
\begin{equation}
U(\theta,\varphi)
:=e^{-i\varphi(I+\sigma_3)/2} e^{-i\theta\sigma_2/2}
=
\begin{pmatrix}
 e^{-i\varphi}\cos (\theta/2) & -e^{-i\varphi}\sin(\theta/2) \\
 \sin(\theta/2) &  \cos(\theta/2) \\
\end{pmatrix}.
\end{equation}
For noninteracting bosons, Bose-Einstein condensation occurs in the mode created by $a_-^\dagger(\bm{0})$.
For interacting bosons, the condensate wave function is gradually modified with increasing the interaction, 
as discussed in the next section. 



\section{Bogoliubov theory and  excitation band topology for homogeneous condensates} \label{sec:Bogoliubov}

In this section, we present the Bogoliubov theory \cite{Pethick08,Pitaevskii03} for homogeneous condensates with weak repulsive interactions $U>0$, 
and determine the band structure of Bogoliubov excitations. 
Here by ``homogeneous'', we refer to the situation in which the system has the periodicity of the honeycomb lattice 
(we do not require the equivalence of the two sublattices). 
We then analyze the topology of the Bogoliubov excitation bands, 
and determine the parameter ranges that exhibit nontrivial topology. 

\subsection{Bogoliubov theory}\label{sec:Bogo}


To formulate the Bogoliubov theory for the present system, 
we first need to determine a condensate wave function by using the Gross-Pitaevskii (GP) theory. 
In the GP theory, we introduce the GP energy functional $E$ by replacing $(a(\rv),a^\dagger(\rv))$ by $(\psi(\rv),\psi^*(\rv))$ in the Hamiltonian \eqref{eq:H}, 
and minimize it with respect to $(\psi(\rv),\psi^*(\rv))$ under the constraint $\sum_\rv |\psi (\rv)|^2=N$. 
Since the single-particle ground state is formed at $\kv=\bm{0}$ (as discussed in Sec.~\ref{sec:band}), 
we introduce the following homogeneous ansatz for the interacting case: 
\begin{equation}\label{eq:psi_X}
\psi (\rv)= \frac{\psi_X}{\sqrt{\Nuc}} ~~ (\rv\in X). 
\end{equation}
We also introduce the chemical potential $\mu$ as a Lagrangian multiplier to satisfy the particle-number constraint. 
The functional to be minimized is then given by
\begin{equation}
E-\mu N= (\psi_A^*, \psi_B^*) [{\cal H} (\bm{0})-\mu I] 
\begin{pmatrix} \psi_A \\ \psi_B \end{pmatrix}
+ \frac{U}{2\Nuc} (|\psi_A|^4+|\psi_B|^4). 
\end{equation}
Minimizing this with respect to $\psi_X^*~(X=A,B)$ gives a homogeneous version of the GP equations:  
\begin{equation}\label{eq:GP_psi_X}
 [{\cal H}(\bm{0}) -\mu I] \begin{pmatrix} \psi_A \\ \psi_B \end{pmatrix}
 +\frac{U}{\Nuc}  \begin{pmatrix} \psi_A^*\psi_A^2 \\ \psi_B^*\psi_B^2 \end{pmatrix} =0.
\end{equation}
Since the single-particle ground state is created by $a_-^\dagger(\bm{0})$ in Eq.~\eqref{eq:aAB_a0pm}, 
it is convenient to parametrize $(\psi_A,\psi_B)^T$ as 
\begin{equation}\label{eq:psiAB}
 \begin{pmatrix} \psi_A \\ \psi_B \end{pmatrix}
 =\sqrt{N} \begin{pmatrix} f_A \\ f_B \end{pmatrix}
 =\sqrt{N} \begin{pmatrix} \sin(\theta/2) \\ \cos(\theta/2) \end{pmatrix},
\end{equation}
where $\theta=\theta_0$ when $U=0$. 
Multiplying Eq.~\eqref{eq:GP_psi_X} by $(f_A,f_B)$ or $(-f_B,f_A)$ from the left, we obtain 
\begin{subequations}\label{eq:mu_theta_eq}
\begin{align}
& h_0(\bm{0})-h(\bm{0})\cos(\theta-\theta_0)+2Un (f_A^4+f_B^4)=\mu, \label{eq:mu_theta}\\
& h(\bm{0})\sin(\theta-\theta_0) + 2Un  f_Af_B(f_A^2-f_B^2)=0, \label{eq:theta_eq}
\end{align}
\end{subequations}
where $n := N/(2\Nuc)$ is the average number of particles per site. 
The parameter $\theta$ is determined by solving Eq.~\eqref{eq:theta_eq}; 
the chemical potential $\mu$ is determined by substituting the obtained $\theta$ into the LHS of \eqref{eq:mu_theta}. 
For the obtained $\theta$, we consider the unitary transformation
\begin{equation}\label{eq:aAB_apm}
 \begin{pmatrix} a_A(\bm{0}) \\ a_B(\bm{0}) \end{pmatrix}
 = U(\theta,\pi) \begin{pmatrix} a_+ \\ a_- \end{pmatrix}
 =\begin{pmatrix} -f_B & f_A \\ f_A & f_B \end{pmatrix}
 \begin{pmatrix} a_+ \\ a_- \end{pmatrix}. 
\end{equation}
Bose-Einstein condensation occurs in the lower-band mode created by $a_-^\dagger$. 
When $\Delta=0$, the two sublattices are equivalent, and thus Eq.~\eqref{eq:theta_eq} gives $\theta=\pi/2$ \cite{Vasic15}. 
For $|\Delta|\ll J_1$, we can expand Eq.~\eqref{eq:mu_theta_eq} in terms of $\theta-\pi/2$, obtaining
\begin{subequations}\label{eq:mu_theta_expand}
\begin{align}
 &\theta=\frac{\pi}{2} -\frac{\Delta}{3J_1+Un} + {\cal O}\left( \Delta^2/J^2 \right), \label{eq:theta_expand}\\
 &\mu=-3J_1+ Un -6J_2\cos\Phi + {\cal O}\left( \Delta^2/J\right).
\end{align}
\end{subequations}
As seen in Eq.~\eqref{eq:theta_expand}, the potential difference $2\Delta$ induces a density imbalance between the two sublattices; 
this imbalance is reduced by a repulsive interaction $U>0$. 


We now discuss excitations from the condensate ground state by using the Bogoliubov theory.  
To this end, using Eqs.~\eqref{eq:ar_ak} and \eqref{eq:aAB_apm}, we decompose $a(\rv)$ into the condensate and noncondensate parts as 
\begin{equation}\label{eq:ar_am_at}
 a(\rv) = \frac{f_X}{\sqrt{\Nuc}} a_- + \tilde{a}(\rv) ~~(\rv\in X).
\end{equation}
Here, the noncondensate part is given by
\begin{equation}\label{eq:at_r}
 \tilde{a}(\rv)
 =\frac{1}{\sqrt{\Nuc}} \bigg[ -\epsilon_X f_{\bar{X}} a_+ + \sum_{\kv\ne\bm{0}} a_X(\kv) e^{i\kv\cdot\rv} \bigg] ~~ (\rv\in X)
\end{equation}
with $\bar{A}=B$ and $\bar{B}=A$. 
Following the Bogoliubov approximation, we replace both $a_-$ and $a^\dagger_-$ by $\sqrt{N}$, 
substitute Eq.~\eqref{eq:ar_am_at} into $H-\mu N$, 
and expand $H-\mu N$ up to quadratic order in $\tilde{a}(\rv)$. 
The terms linear in $\tilde{a}(\rv)$ or $\tilde{a}^\dagger(\rv)$ disappear because of the stability condition of the condensate in Eq.~\eqref{eq:theta_eq}, 
and we arrive at the Bogoliubov Hamiltonian
\begin{equation}\label{eq:H_bogo}
H -\mu N=\frac12 \sum_{\kv\ne\bm{0}} \alphav^\dagger(\kv) {\cal M}(\kv) \alphav(\kv)
+ \frac12 \left(a_+^\dagger,a_+\right) {\cal M}_+
 \begin{pmatrix} a_+ \\ a_+^\dagger \end{pmatrix}
 +\mathrm{const.}
\end{equation}
with
\begin{equation}
 \alphav^\dagger (\kv) := \left( a_A^\dagger(\kv), a_B^\dagger(\kv), a_A(-\kv), a_B(-\kv) \right).
\end{equation}
Here, we have introduced the $4\times 4$ matrix ${\cal M}(\kv)$ and the $2\times 2$ matrix ${\cal M}_+$ as
\begin{align}
&{\cal M}(\kv) = 
\begin{pmatrix}
{\cal H}(\kv) -\mu I + 4Un  F^2 & 2Un F^2 \\
2Un F^2 & {\cal H}^T(-\kv) -\mu I + 4Un F^2
\end{pmatrix}, \label{eq:M_k}\\
&{\cal M}_+ =
 \left[ h_0(\bm{0}) + h(\bm{0}) \cos(\theta-\theta_0) -\mu + 8Un f_A^2f_B^2 \right] I
+ 4Un f_A^2 f_B^2 \sigma_1
\end{align}
with $F:=\mathrm{diag}(f_A,f_B)$.


To diagonalize the Bogoliubov Hamiltonian \eqref{eq:H_bogo}, 
we perform generalized Bogoliubov transformations
\begin{equation}
 \alphav(\kv)=W(\kv) \betav(\kv) ,~
 \begin{pmatrix} a_+ \\ a_+^\dagger \end{pmatrix}
 = W_+
 \begin{pmatrix} b_+(\bm{0}) \\ b_+^\dagger (\bm{0}) \end{pmatrix}
\end{equation}
with
\begin{equation}
 \betav^\dagger (\kv) := \left( b_+^\dagger(\kv), b_-^\dagger(\kv), b_+(-\kv), b_-(-\kv) \right).
\end{equation}
Here,  $W(\kv)$ and $W_+$ are paraunitary matrices which satisfy
\begin{equation}\label{eq:WtW}
 W^\dagger (\kv) \tau_3 W(\kv) = W(\kv) \tau_3 W^\dagger (\kv) = \tau_3 ,~
 W_+^\dagger \sigma_3 W_+ = W_+ \sigma_3 W_+^\dagger = \sigma_3
\end{equation}
with $\tau_3:= \mathrm{diag}(1,1,-1,-1)$.  
These equations ensure the invariance of the bosonic commutation relations. 
If the matrices  $W(\kv)$ and $W_+$ are chosen to satisfy 
\begin{equation}\label{eq:WMW}
\begin{split}
 &W^\dagger (\kv) {\cal M}(\kv) W(\kv)
 = \mathrm{diag}(E_+(\kv),E_-(\kv),E_+(-\kv),E_-(-\kv)),\\\
 &W_+^\dagger {\cal M}_+ W_+= E_+(\bm{0}) I,
\end{split}
\end{equation}
the Bogoliubov Hamiltonian \eqref{eq:H_bogo} is diagonalized as
\begin{equation}\label{eq:H_bb}
 H-\mu N = \sum_{\kv} E_+(\kv) b_+^\dagger(\kv) b_+(\kv) 
 +\sum_{\kv\ne\bm{0}} E_-(\kv) b_-^\dagger(\kv) b_-(\kv)+\mathrm{const.} .
\end{equation}
The paraunitary matrix $W(\kv)$ satisfying Eq.~\eqref{eq:WMW} 
can be constructed numerically by using the method described in Refs.~\cite{Shindou13,Colpa78}. 
Examples of the calculated Bogoliubov excitation bands $E_\pm(\kv)$ are presented in Fig.~\ref{fig:bands}(b,c). 

\subsection{Band gap}\label{sec:bandgap}

\begin{figure}
\begin{center}\includegraphics[width=0.6\textwidth]{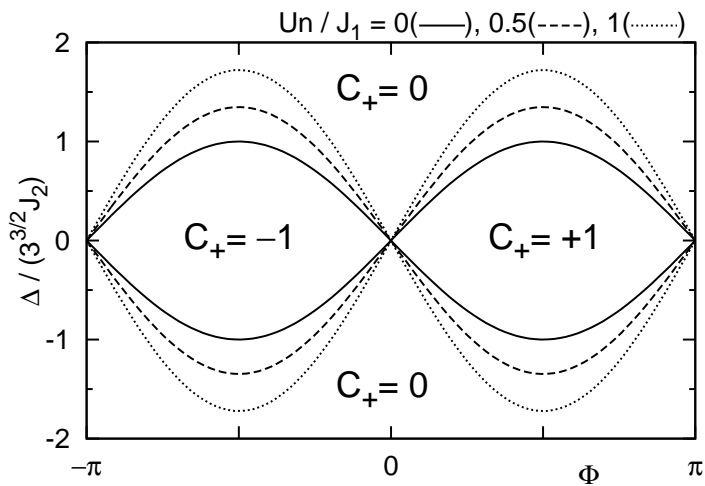}
\cutspace
\caption{\label{fig:phase}
The Chern number $C_+$ of the higher excitation band in the Haldane-Bose-Hubbard model with $J_2,|\Delta|\ll J_1$.
The boundaries between regions of different $C_+$ values are given by Eq.~\eqref{eq:phase_HBH}, and indicated for $Un/J_1=0,0.5,1$. 
The solid curves correspond to the noninteracting case \cite{Haldane88}. 
The regions with nontrivial topology ($C_+\ne 0$) enlarge as $Un/J_1$ increases. 
}\end{center}
\end{figure}

Before discussing the topology of the Bogoliubov excitation bands, 
let us analyze the gap between the two bands $E_\pm(\kv)$. 
The band topology cannot change as far as the band gap remains open. 
Thus the closing of the gap can signal a change in topology. 

As seen in Fig.~\ref{fig:bands} and known in the noninteracting case \cite{Haldane88}, 
the smallest gap is found at one of the two points 
$\kv=\pm \Kv:=\pm\left( \av_1^*+\av_2^* \right)/3$
in the Brillouin zone; see the $K_\pm$ points in Fig.~\ref{fig:model}. 
At these points, because of $h_1(\pm\Kv)=h_2(\pm\Kv)=0$, 
the matrix ${\cal M}(\pm\Kv)$ in Eq.~\eqref{eq:M_k} is decoupled into $A$ and $B$ sublattice blocks, 
each of which can be diagonalized by a standard $2\times 2$ Bogoliubov transformation. 
The excitation energies at $\kv=\pm\Kv$ are then calculated to be
\begin{align}
 &E_X(\pm\Kv) = \lambda_X\pm 3\sqrt{3}\epsilon_X J_2\sin\Phi ~~(X=A,B), \label{eq:EXpmK}\\
 &\lambda_X:=\left[ (3J_2\cos\Phi+\epsilon_X\Delta-\mu+4Un f_X^2)^2-(2Un f_X^2)^2 \right]^{1/2} \label{eq:lambdaX}. 
\end{align}
The higher (lower) of these energies gives $E_+(\pm\Kv)$ ($E_-(\pm\Kv)$) in Eq.~\eqref{eq:H_bb}. 
The gap-closing conditions at $\kv=\pm\Kv$ are thus obtained as 
\begin{align}\label{eq:phase_HBH_exact}
0=E_A(\pm\Kv)-E_B(\pm\Kv)= \lambda_A-\lambda_B \pm 6\sqrt{3} J_2\sin\Phi.
\end{align}
For $J_2,|\Delta|\ll J_1$, by using \eqref{eq:mu_theta_expand}, we can expand $\lambda_A-\lambda_B$ as
\begin{equation}
\lambda_A-\lambda_B= \frac{2\Delta}{G(Un /J_1)} + {\cal O} \left( \frac{J_2^2}{J_1},\frac{\Delta^2}{J_1}\right),~
G(s):=\left(1+\frac{2s}{3}\right)^{\frac12}\left(1+\frac{s}{3}\right). 
\end{equation}
Equation \eref{eq:phase_HBH_exact} is then rewritten into a simple form
\begin{equation}\label{eq:phase_HBH}
 \Delta \pm 3\sqrt{3}J_2 G(Un /J_1) \sin\Phi + {\cal O} \left( J_2^2/J_1,\Delta^2/J_1\right)=0.
\end{equation}
For $Un=0$, this reduces to the exact phase boundaries $\Delta=\mp 3\sqrt{3}J_2\sin\Phi$ in the noninteracting case \cite{Haldane88}. 
Equation~\eqref{eq:phase_HBH} indicates that with increasing the strength of interaction $Un/J_1$, 
these boundaries are shifted to larger $|\Delta|$ by a factor of $G(Un/J_1)$; see Fig.~\ref{fig:phase}. 
Here, $G(s)$ is an increasing function of $s$, and expanded for $|s|\ll 1$ as 
\begin{equation}
 G(s)=1+ 2s/3+s^2/18+{\cal O}(s^4).
\end{equation}
Across the gap-closing lines, the topology of the Bogoliubov excitation bands changes, as we discuss next. 

\subsection{Chern number} \label{sec:chern}


We now analyze the topology of the Bogoliubov excitation bands. 
We first note that technically, ${\cal M}(\kv)$ in Eq.~\eqref{eq:M_k} (more specifically, ${\cal H}(\kv)$ in it) needs a modification for such a purpose  
because it is {\it not} periodic in the first Brillouin zone in the present representation. 
This is because the Fourier expansion \eqref{eq:ar_ak} was based on the real-space positions $\rv$ 
--- while this treatment was useful in making ${\cal H}(\kv)$ possess the $C_3$ symmetry of the original lattice, 
the spacing between the two sublattices introduced an additional phase factor, 
which broke the periodicity in the first Brillouin zone. 
To recover the periodicity, we define $\tilde{\cal M}(\kv)$ by replacing ${\cal H}(\kv)$ by 
$\tilde{\cal H}(\kv) = e^{-i\kv\cdot\deltav_3(I-\sigma_3)/2} {\cal H}(\kv) e^{i\kv\cdot\deltav_3(I-\sigma_3)/2}$ in Eq.~\eqref{eq:M_k}. 
We then introduce the paraunitary matrix $\tilde{W}(\kv)$ as the matrix which ``diagonalizes'' $\tilde{\cal M}(\kv)$ in the sense of Eq.~\eqref{eq:WMW}.

The $4\times 4$ paraunitary matrix $\tilde{W}(\kv)$ can be parametrized as
\begin{equation}
 \tilde{W}(\kv) = 
\begin{pmatrix}
 {\cal U}(\kv) & {\cal V}^*(-\kv) \\
 {\cal V}(\kv) & {\cal U}^*(-\kv)
\end{pmatrix}
\end{equation}
with 
\begin{equation}
{\cal U}(\kv)=
\begin{pmatrix}
 u_{A+}(\kv) & u_{A-}(\kv) \\
 u_{B+}(\kv) & u_{B-}(\kv) 
\end{pmatrix}, ~
{\cal V}(\kv)=
\begin{pmatrix}
v_{A+}(\kv) & v_{A-}(\kv) \\
v_{B+}(\kv) & v_{B-}(\kv)
\end{pmatrix}.
\end{equation}
To discuss the topology of each excitation band, we introduce the vectors
\begin{equation}
 |w_\gamma(\kv)\rangle = (u_{A\gamma}(\kv),u_{B\gamma}(\kv),v_{A\gamma}(\kv),v_{B\gamma}(\kv))^T~~(\gamma=\pm), 
\end{equation}
where $\gamma=+$ and $-$ correspond respectively to the first and second columns of $W(\kv)$.
It follows from Eqs.~\eqref{eq:WtW} and \eqref{eq:WMW} that 
these vectors satisfy the eigen equation 
\begin{equation}\label{eq:MwEtw}
 \tilde{\cal M} (\kv) |w_\gamma(\kv)\rangle = E_\gamma (\kv) \tau_3 |w_\gamma(\kv)\rangle 
\end{equation}
and the orthonormality condition
\begin{equation}
 \langle w_\gamma(\kv)| \tau_3 |w_{\gamma'}(\kv)\rangle =\delta_{\gamma\gamma'}. 
\end{equation}


For each band, we introduce the Berry curvature \cite{Haldane08,Shindou13}
\begin{equation}
 {\cal B}_\gamma (\kv) = i\epsilon_{ij} \langle \partial_i w_\gamma(\kv) | \tau_3 |\partial_j w_\gamma(\kv) \rangle
\end{equation}
with $\partial_i:=\frac{\partial}{\partial k_i}$. 
The Chern number can then be introduced as \cite{Thouless82,Haldane08,Shindou13}
\begin{equation}
 C_\gamma = \int_\mathrm{BZ} \frac{d^2\kv}{2\pi} {\cal B}_\gamma(\kv).  
\end{equation}
In the noninteracting case $Un=0$, both $C_\pm$ are quantized to integers, 
and satisfy the zero sum rule $\sum_\gamma C_\gamma=0$. 
In the interacting case $Un>0$, however, $|w_-(\kv)\rangle$ is {\it not} defined at $\kv=\bm{0}$,\footnote{
This is because the Bogoliubov excitations consist only of modes orthogonal to the GP ground state \cite{Pethick08,Pitaevskii03}; see Eq.~\eqref{eq:at_r}.
} 
and thus there is an ambiguity in the definition of $C_-$. 
Nevertheless, $C_+$ is still well-defined, and quantized to an integer.\footnote{
This does not contradict the zero sum rule. In the interacting case $Un>0$, 
the rule applies to the sum over all the particle and hole bands. 
Namely, $C_++C_-+C_+'+C_-'=0$, where $C_\pm'$ are the Chern numbers associated with the hole bands \cite{Shindou13}. 
One can easily show $C_++C_+'=0=C_-+C_-'$, and thus the sum rule is trivially satisfied. 
Therefore, one cannot use the ill-defined nature of $C_-$ to change $C_+$ to an arbitrary value. 
} 
We numerically calculate $C_+$ using the manifestly gauge-invariant method proposed in Ref.~\cite{Fukui05}. 

The ``phase diagram'' based on the Chern number $C_+$ is presented in Fig.~\ref{fig:phase} 
(we note that this diagram is {\it not} based on ground-state transitions). 
We numerically confirmed that the boundaries between regions of different $C_+$ values are given precisely 
by the gap-closing condition \eqref{eq:phase_HBH_exact} (or Eq.~\eqref{eq:phase_HBH} for $J_2,|\Delta|\ll J_1$) obtained in Sec.~\ref{sec:bandgap}.  
The obtained results indicate that the topology of the Bloch bands known for the noninteracting case $Un=0$ \cite{Haldane88} 
are smoothly carried over to that of the Bogoliubov excitation bands for the interacting case $Un>0$, 
and that the regions displaying nontrivial topology $C_+\ne 0$ enlarge  with increasing $Un/J_1$. 
When $C_+\ne 0$, the bulk-edge correspondence \cite{Hatsugai93} dictates that in-gap chiral edge modes intervening between the upper and lower bulk bands 
appear when the system has a boundary. 
We numerically demonstrate the emergence of such modes in Sec.~\ref{sec:inhomo_BEC_numerics}. 

In closing this section, two remarks are in order.  

The first remark is on the reason why the ranges displaying topological bands expand with increasing the interaction $U$. 
In the noninteracting case $U=0$, the potential difference $2\Delta$ between the two sublattices 
drives a transition from topological to trivial bands 
as it favors sublattice-separated Bloch wave functions, which have trivial topology. 
Algebraically, this potential difference induces a finite difference between $\lambda_A$ and $\lambda_B$ defined in Eq.~\eqref{eq:lambdaX}, 
and the transition occurs when $|\lambda_A-\lambda_B|=6\sqrt{3}J_2 |\sin\Phi|$. 
The interaction $U>0$ has the effect of obscuring this difference [through $Unf_X^2$ in Eq.~\eqref{eq:lambdaX}], 
and thus a larger $|\Delta|$ is required to drive the topological-to-trivial transition. 
It will be interesting to investigate whether a similar stabilization of topological bands occurs in a wider variety of interacting systems. 

The second remark is on the case of attractive interactions $U<0$. 
While an attractive Bose gas is unstable against collapse in the thermodynamic limit, 
it can form a metastable condensate in a finite system if $N$ is below a certain critical value \cite{Ruprecht95}. 
As far as a quasi-homogeneous condensate is realized, 
we can perform the same Bogoliubov analysis as in this section, 
and obtain the phase boundaries of topologically nontrivial regions as in Eq.~\eqref{eq:phase_HBH}; 
these regions gradually shrink with increasing $|U|n$ for $U<0$. 
In these regions, in-gap chiral edge modes discussed in Sec.~\ref{sec:inhomo_BEC_numerics} are also expected to be formed. 

\section{Ground state and excitations in trapped condensates: formalism}\label{sec:inhomo_BEC_formalism}

In this section, we present the formalism for calculating the ground state and excitations of trapped condensates. 
We first describe the Bogoliubov-de Gennes (BdG) theory \cite{Pethick08,Pitaevskii03} for inhomogneous condensates on lattices. 
We then apply this theory to a strip geometry, which is convenient for discussing edge modes. 
We also describe an extended Thomas-Fermi approximation 
which can give a simple analytic expression for the density profile in a given trap potential. 
Numerical results obtained using the formalism are presented in Sec.~\ref{sec:inhomo_BEC_numerics}. 

\subsection{Bogoliubov-de Gennes theory for inhomogeneous condensates}


The BdG theory for inhomogeneous condensates can be derived by linearizing a time-dependent GP equation. 
We start from the Heisenberg equation of motion for the time-dependent operator $a(\rv,t)$: 
\begin{equation}
 i\hbar \partial_t a(\rv,t)
 = [a(\rv,t), H] 
 = -\sum_{\rv'} J(\rv,\rv') a(\rv',t) + U a^\dagger(\rv,t) a(\rv,t)^2.
\end{equation}
Replacing $(a(\rv,t),a^\dagger(\rv,t))$ by classical fields $(\psi(\rv,t),\psi^*(\rv,t))$, we obtain the GP equation
\begin{equation}\label{eq:GP}
 i \hbar \partial_t \psi(\rv,t) = - \sum_{\rv'} J(\rv,\rv') \psi(\rv',t) + U\psi^*(\rv,t) \psi(\rv,t)^2. 
\end{equation}
Equation \eqref{eq:particle_N} imposes the normalization condition $\sum_\rv |\psi (\rv,t)|^2=N$. 
Inserting a stationary state ansatz $\psi(\rv,t)=\psi(\rv) e^{-i\mu t/\hbar}$ into Eq.~\eqref{eq:GP}, we obtain the time-independent GP equation 
\begin{equation}\label{eq:GP_r}
 \mu \psi (\rv) = - \sum_{\rv'} J (\rv,\rv') \psi(\rv') + U \psi^*(\rv) \psi(\rv)^2.
\end{equation}
The solution $\psi(\rv)$ with the lowest frequency $\mu/\hbar$ gives the GP ground state. 


We now discuss small fluctuations around the GP ground state: 
$\psi (\rv,t)=\psi(\rv) e^{-i\mu t/\hbar}+\phi(\rv,t)$. 
Expanding the GP equation \eqref{eq:GP} to first order in $\phi(\rv)$, we obtain a linear differential equation
\begin{equation}\label{eq:GP_linear}
\begin{split}
 i\hbar \partial_t \phi(\rv,t) 
 = &- \sum_{\rv'} J(\rv,\rv') \phi(\rv',t) 
 + 2U |\psi(\rv)|^2 \phi(\rv,t) 
 + U e^{-2i\mu t/\hbar} \psi(\rv)^2 \phi^*(\rv,t).
\end{split}
\end{equation}
Assuming a solution of the form 
\begin{equation}
 \phi(\rv,t)= e^{-i\mu t/\hbar}  \left[u(\rv) e^{-iEt/\hbar} + v^*(\rv) e^{iEt/\hbar} \right],
\end{equation}
we obtain Bogoliubov-de Gennes (BdG) equations 
\begin{subequations}\label{eq:BdG}
\begin{align}
 +E u(\rv)  =& -\sum_{\rv'} J(\rv,\rv') u(\rv') + (2U|\psi(\rv)|^2-\mu) u(\rv) + U\psi(\rv)^2 v(\rv),\\
 -E v(\rv) =&  -\sum_{\rv'} J^*(\rv,\rv') v(\rv') + (2U|\psi(\rv)|^2-\mu) v(\rv) + U\psi^*(\rv)^2 u(\rv).
\end{align}
\end{subequations}
If the condensate is stable, these equations admit $N_s-1$ (linearly independent) sets of solutions $(u_j(\rv),v_j(\rv))$ with positive frequencies $E_j/\hbar$, 
where $N_s$ is the total number of lattice sites. 
Such solutions can be chosen to satisfy the orthonormality condition
\begin{equation}
 \sum_\rv [u_j^*(\rv) u_l(\rv)-v_j^*(\rv) v_l(\rv)] =\delta_{jl}.
\end{equation}


The present formulation of the BdG theory is based on the linearization of the GP equation \eqref{eq:GP}, 
which is an equation of motion for the classical field. 
However, we can check the consistency of this classical-field formulation with the operator formulation for the homogeneous case in Sec.~\ref{sec:Bogoliubov}. 
Indeed, substituting the ansatz \eqref{eq:psi_X} into the GP equation \eqref{eq:GP_r} reproduces Eq.~\eqref{eq:GP_psi_X}. 
Furthermore, substituting 
\begin{equation}
 u(\rv)=\frac{u_{X\gamma}(\kv)}{\sqrt{\Nuc}} e^{i\kv\cdot\rv},~
 v(\rv)=\frac{v_{X\gamma}(\kv)}{\sqrt{\Nuc}} e^{i\kv\cdot(\rv-\frac{1-\epsilon_X}{2} \deltav_3)}~~(\rv\in X)
\end{equation}
into the BdG equation \eqref{eq:BdG} reproduces Eq.~\eqref{eq:MwEtw}.
It is known that the operator formulation for the inhomogeneous case 
also leads to the same set of BdG equations as in Eq.~\eqref{eq:BdG} \cite{Pethick08,Pitaevskii03}. 

\subsection{Strip geometry}\label{sec:strip}

\begin{figure}
\begin{center}\includegraphics[width=0.6\textwidth]{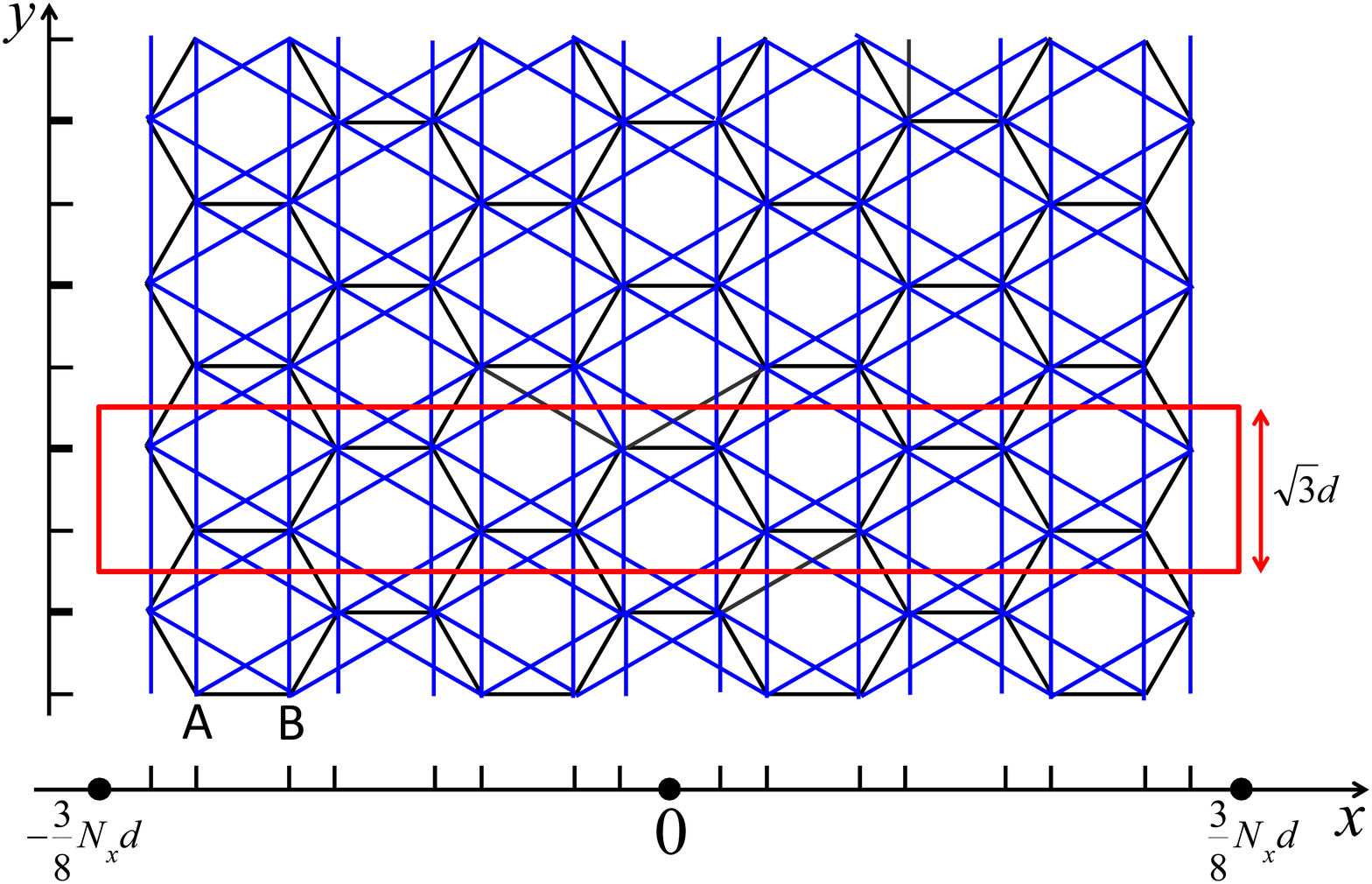}
\cutspace
\caption{\label{fig:strip}
Strip (or ``nanoribon'') geometry with zigzag edges. 
A periodic boundary condition is imposed only along the $y$ direction. 
The red square indicates a unit cell which contains $N_x=16$ lattice sites. 
We denote the number of unit cells by $N_y$. 
The origin of the $x$ coordinate is placed at the center of the system. 
}\end{center}
\end{figure}


We apply the BdG theory to analyze excitations in trapped condensates. 
We describe site positions by two-dimensional coordinates $\rv=(x,y)$. 
For simplicity, we consider a strip geometry, which is periodic only along the $y$ direction as in Fig.~\ref{fig:strip} and analogous to a graphene nanoribbon. 
This geometry can be used to describe the central part of an elongated condensate 
around which the system is approximately uniform in the elongated direction. 
Along the $x$ direction, we introduce a box trap with sharp zigzag edges (Fig.~\ref{fig:strip}) or a harmonic trap \cite{Buchhould12} in Sec.~\ref{sec:inhomo_BEC_numerics}; 
however, we do not assume a specific trap potential in this subsection. 


Exploiting the translation invariance along the $y$ direction, 
we make the following ansatz for the GP ground state: 
\begin{equation}\label{eq:psi_xy}
 \psi(x,y)=\frac{\psi_x}{\sqrt{N_y}}=\sqrt{n_y} f_x,~ n_y:=\frac{N}{N_y},
\end{equation}
where $N_y$ is the number of unit cells, 
and $\psi_x=\sqrt{N}f_x$ satisfies the normalization condition $\sum_x |\psi_x|^2=N\sum_x|f_x|^2=N$.
Substituting this into the GP equation \eqref{eq:GP_r}, we obtain
\begin{equation}\label{eq:GP_fx}
 \mu f_x = \sum_{x'} {\cal H}_{xx'}(0) f_{x'} + Un_y f_x^* f_x^2. 
\end{equation}
Here we have introduced 
\begin{equation}\label{eq:H_xx_ky}
 {\cal H}_{xx'} (k_y) = -\sum_{y} J(x,y,x',y') e^{-ik_y(y-y')} ,
\end{equation}
where $(x',y')$ is a particular site position, and the sum over $y$ is restricted to the site positions for fixed $x$; 
the translation invariance along the $y$ direction ensures that the RHS of Eq.~\eqref{eq:H_xx_ky} does not depend on $y'$.  
An accurate solution to Eq.~\eqref{eq:GP_fx} with the lowest frequency $\mu/\hbar$ 
can be obtained by numerically performing the imaginary time evolution with the replacement $\mu\to - \partial_\tau$. 


After obtaining the GP ground state \eqref{eq:psi_xy}, 
we solve the BdG equations \eqref{eq:BdG} to calculate excitations. 
We introduce the following ansatz with momentum $k_y$ in the $y$ direction: 
\begin{equation}
 u(x,y)=\frac{u_x}{\sqrt{N_y}} e^{ik_yy},~
 v(x,y)=\frac{v_x}{\sqrt{N_y}} e^{ik_yy}. 
\end{equation}
Substituting these into the BdG equation \eqref{eq:BdG}, we obtain an eigen equation
\begin{equation}\label{eq:Mky_eig}
 {\cal M} (k_y) \wv = E \tau_3 \wv. 
\end{equation}
Here we have introduced a $2N_x$-component vector $\wv=(\{u_x\},\{v_x\})^T$, 
and a $2N_x\times 2N_x$ matrix
\begin{equation}
{\cal M} (k_y) 
= 
\begin{pmatrix}
{\cal H}(k_y) -\mu I + 2Un_y F^*F & Un_y F^2 \\
Un_y F^{*2} & {\cal H}^T(-k_y) -\mu I + 2Un_y F^*F
\end{pmatrix},\\
\end{equation}
with $F:=\mathrm{diag}(\{f_x\})$. 
The eigen equation \eqref{eq:Mky_eig} can be solved numerically by using the method of Ref.~\cite{Colpa78}. 
We denote positive-frequency solutions by $\wv_j(k_y)=(\{u_{xj}(k_y)\},\{v_{xj}(k_y)\})^T$ and $E_j(k_y)$ with $j=1,2,\dots,N_x$ in descending order in energy. 
In the operator formulation of the BdG theory, this leads to the fact that 
the Hamiltonian is diagonalized as
\begin{equation}\label{eq:H_BdG}
 H-\mu N = \sum_{(k_y,j)\ne (0,N_x)} E_j(k_y) b_j^\dagger(k_y) b_j(k_y) + \mathrm{const.},
\end{equation}
where the new bosonic operators $\{b_j(k_y)\}$ are related to the original ones as
\begin{equation}\label{eq:a_psi_b}
 a(x,y) = \psi(x,y)+\frac{1}{\sqrt{N_y}} \sum_{(k_y,j)\ne (0,N_x)} e^{ik_y y} \left[ u_{xj}(k_y) b_j(k_y) + v_{xj}^*(-k_y) b_j^\dagger(-k_y) \right].
\end{equation}


To discuss excitations in a strip geometry, 
it is useful to introduce the spectral weight at zero temperature 
\begin{equation}\label{eq:rho_xxky}
 \rho (x,x',k_y,\omega)=\int_{-\infty}^\infty dt \sum_y e^{-ik_y (y-y')+i\omega t} \langle [a(x,y,t),a^\dagger(x',y',0)] \rangle,
\end{equation}
where $(x,y)$ and $(x',y')$ are taken similarly as in Eq.~\eqref{eq:H_xx_ky}, 
and the average $\langle\cdot\rangle$ is taken over the vacuum of the Bogoliubov excitations in Eq.~\eqref{eq:H_BdG}. 
Using Eqs.~\eqref{eq:H_BdG} and \eqref{eq:a_psi_b}, Eq.~\eqref{eq:rho_xxky} can be rewritten as
\begin{equation}\label{eq:rho_xxky_calc}
\begin{split}
 \rho(x,x',k_y,\omega) = 2\pi \sum_j 
&\big[ u_{xj}(k_y)u_{x'j}^*(k_y) \delta(\omega-E_j(k_y)/\hbar) \\
&+ v_{xj}^*(-k_y)v_{x'j}(-k_y) \delta(\omega+E_j(-k_y)/\hbar)\big].
\end{split}
\end{equation}
In Sec.~\ref{sec:inhomo_BEC_numerics}, we calculate $\rho(x,x,k_y,\omega)$ to discuss excitations that can be probed at each position $x$. 
We note that similar calculations are performed for a fermionic Hofstadter model in various traps by Buchhould {\it et al.} \cite{Buchhould12}. 

\subsection{Extended Thomas-Fermi approximation for density profiles}

\newcommand{\nTF}{n_\mathrm{TF}^\mathrm{max}}
\newcommand{\RTF}{R_\mathrm{TF}}


Here we provide an approximate solution to the scaled GP equation \eqref{eq:GP_fx}. 
This helps us tune the potential and the interaction to obtain the desired density profiles later. 
We extend the Thomas-Fermi approximation \cite{Pethick08,Pitaevskii03}, 
where the condensate wave function is assumed to vary slowly in space,  
so that the oscillations of $f_x$ between the two sublattices due to $\Delta\ne 0$ are also taken into account. 
We assume
\begin{equation}
 f_x = \bar{f}(x)+ \epsilon_x \delta f(x), 
\end{equation}
where $\bar{f}(x)$ and $\delta f(x)$ are slowly varying real functions, 
and $\epsilon_x=+1$ ($-1$) if $x$ belongs to the $A$ ($B$) sublattice. 
As can be seen in Fig.~\ref{fig:strip}, all the sites having the same $x$ belong to the same sublattice. 
Assuming $|\Delta|\ll J_1$, we can expect $|\delta f(x)| \ll \bar{f}(x)$.
Then the RHS of Eq.~\eqref{eq:GP_fx} can be approximated as
\begin{equation}
\begin{split}
 &\left[ \sum_{x'=x-3/2}^{x+3/2} {\cal H}_{xx'}(0) \right] \bar{f}(x) 
 + \left[ \sum_{x'=x-3/2}^{x+3/2} \epsilon_{x'} {\cal H}_{xx'}(0) \right] \delta f(x) + Un_y [\bar{f}(x)+\delta f(x)]^3 \\
 &\approx \left[ V(x) -3J_1-6J_2\cos\Phi +Un_y \bar{f}(x)^2 \right] \bar{f} (x)\\
 &~~~+ \epsilon_x \left[ \Delta \bar{f}(x) + (V(x)+3J_1-6J_2\cos\Phi+3Un_y \bar{f}(x)^2) \delta f(x) \right],
\end{split}
\end{equation}
where we have ignored higher-order terms in $\Delta$ and $\delta f(x)$. 
Here, the first and second lines of the last expression can be viewed as uniform and staggered components  
because $\epsilon_x$ oscillates rapidly and other functions of $x$ vary slowly. 
Requiring that Eq.~\eqref{eq:GP_fx} holds separately for different components (because they cannot cancel each other), we obtain
\begin{subequations}
\begin{align}
 &Un_y \bar{f}(x)^2 = \max (U\nTF-V(x), 0),~~U\nTF :=\mu+3J_1+6J_2\cos\Phi,\\
 &\delta f(x)=-\frac{\Delta \bar{f}(x)}{V(x)-\mu+3J_1-6J_2\cos\Phi+3Un_y\bar{f}(x)^2}. 
\end{align}
\end{subequations}
Therefore, the density profile scaled by the interaction $U$ is obtained as
\begin{equation}\label{eq:Unfx2_TF}
 Un_y |f_x|^2 \approx \max (U\nTF-V(x), 0) \left( 1- \frac{\epsilon_x \Delta}{U\nTF-V(x)+3J_1}\right).
\end{equation}
This expression agrees well with the density profiles of the numerically calculated GP ground states presented in Sec.~\ref{sec:inhomo_BEC_numerics}. 
As seen in this expression, $\nTF$ can be interpreted as the maximal uniform-component density achieved at the potential minimum with $V(x)=0$. 

\section{Ground state and excitations in trapped condensates: numerical results}\label{sec:inhomo_BEC_numerics}

In this section, we present numerical results on the ground state and excitations of trapped condensates 
based on the formalism described in Sec.~\ref{sec:inhomo_BEC_formalism}. 
We exploit the strip geometry described in Sec.~\ref{sec:strip}, 
and introduce a box trap or a harmonic trap along the $x$ direction. 
While harmonic traps are used more commonly in experiments of ultracold atoms, 
a box trap with sharp boundaries has also been realized recently \cite{Gaunt13}. 
Sharp boundaries can also be realized in synthetic dimensions \cite{Celi14,Mancini15,Stuhl15}. 
We demonstrate that chiral edge states reflecting the nontrivial bulk band topology do appear in a box trap. 
For a harmonic trap, by contrast, our results show that such edge states are substantially obscured and difficult to observe, 
as opposed to an expectation from a semiclassical picture. 

\subsection{Box trap}


We first consider the case of a box trap 
\begin{equation}\label{eq:V_box}
 V(x)=
\begin{cases}
 0 & \left(|x|<\frac38 N_x d\right);\\
 \infty & (\mathrm{otherwise}),
\end{cases}
\end{equation}
which has sharp boundaries of zigzag type as in Fig.~\ref{fig:strip}. 
The extended TF result \eqref{eq:Unfx2_TF} leads to a uniform density in each sublattice inside the box. 
Summing Eq.~\eqref{eq:Unfx2_TF} over the lattice sites $|x|<\frac38 N_x d$, we find $Un_y \approx U \nTF N_x$. 
We note that in the BdG calculation for a strip geometry, the product $Un_y$ is an input parameter, 
i.e., if $Un_y$ is fixed, the result does not depend on individual values of $U$ and $n_y$. 
We can thus tune $Un_y$ to obtain a desired scaled average density $U\nTF /J_1 \approx Un_y/(N_xJ_1)$. 

\begin{figure}
\begin{center}\includegraphics[width=0.6\textwidth]{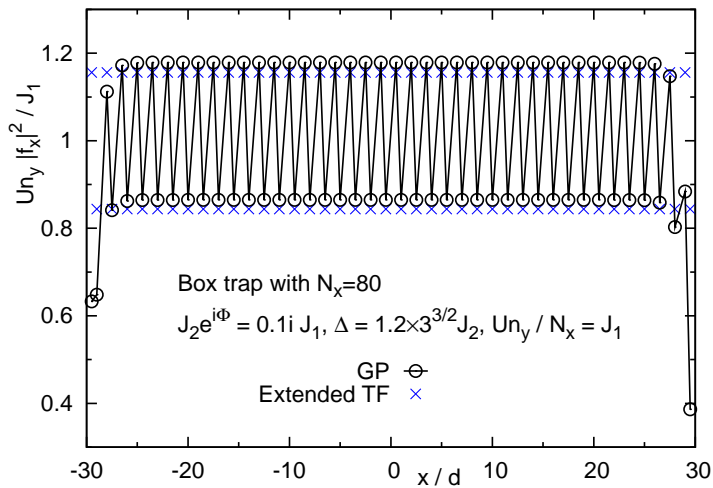}
\cutspace
\caption{\label{fig:gs_strip}
Density profile (scaled by $U/J_1$) of the ground state along the $x$ direction 
in a box trap \eqref{eq:V_box} with $N_x=80$ (see Fig.~\ref{fig:gs_strip}). 
The data for the GP ground state are compared with the extended TF result \eqref{eq:Unfx2_TF}. 
The model parameters are chosen such that the case of Fig.~\ref{fig:bands}(c) 
with interaction-induced nontrivial topology $C_+=+1$ 
is realized in a quasi-homogeneous region in the bulk. 
}\end{center}
\end{figure}


We set $(J_2e^{i\Phi}/J_1,\Delta/J_2,Un_y/(N_xJ_1))=(0.1i,1.2\times 3^{3/2},1)$  
so that the case of Fig.~\ref{fig:bands}(c) is realized in a quasi-homogeneous region in the bulk. 
The scaled density profile $Un_y |f_x|^2 / J_1$ of the GP ground state in Fig.~\ref{fig:gs_strip}
is indeed almost uniform in each sublattice, and agrees well with the extended TF result \eqref{eq:Unfx2_TF}, 
except over a few sites near each boundary. 

\begin{figure}
\begin{center}\includegraphics[width=0.9\textwidth]{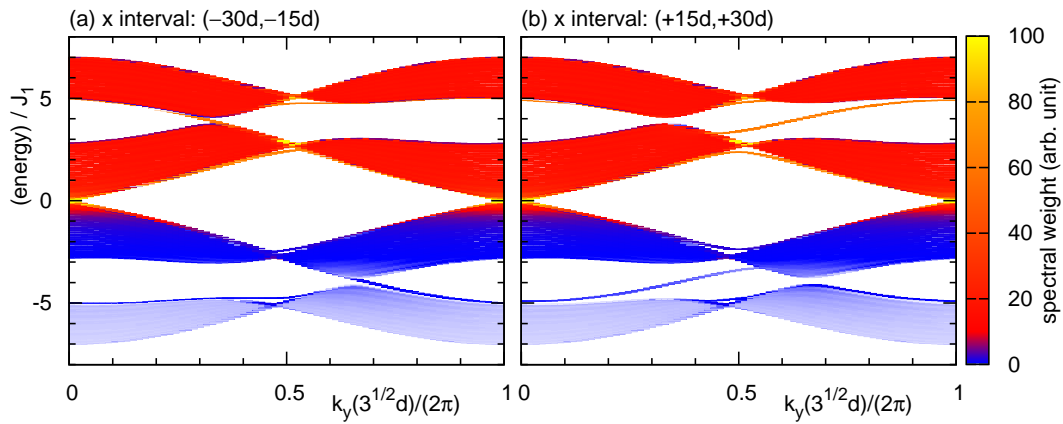}
\cutspace
\caption{\label{fig:spec_strip}
Integrated spectral weight $\sum_{x\in I} \rho(x,x,k_y,\omega)$ for  
(a) $x$ in $I=(-30d,-15d)$ and (b) $x$ in $I=(+15d,+30d)$, which contain the left and right edges, respectively. 
The BdG calculation is performed using the GP ground state shown in Fig.~\ref{fig:gs_strip}, 
and the spectral weight $\rho(x,x,k_y,\omega)$ is calculated with Eq.~\eqref{eq:rho_xxky}. 
Positive (negative) energies correspond to particle (hole) excitations.  
}\end{center}
\end{figure}


We now discuss excitations calculated by the BdG theory described in Sec~\ref{sec:strip}. 
For an interval $I$ of the $x$ coordinate, we consider the integrated spectral weight $\sum_{x\in I} \rho(x,x,k_y,\omega)$, 
where $\rho(x,x,k_y,\omega)$ is defined in Eq.~\eqref{eq:rho_xxky}. 
The results for (a) $I=(-30d,-15d)$ and (b) $I=(+15d,+30d)$ are presented in Fig.~\ref{fig:spec_strip}. 
In both of these results, the continuum of excitations corresponding to the two bulk particle (hole) excitation bands 
is found with high (low) spectral density. 
Furthermore, inside the band gaps, chiral modes with negative and positive velocities are clearly formed in (a) and (b), respectively.\footnote{
In Fig.~\ref{fig:spec_strip} [and Fig.~\ref{fig:bands}(c)], the bulk band gap is relatively small 
because we examine the case in which nontrivial topology $C_+=+1$ is induced by the interaction $U$. 
That is, the system is located in a narrow region between the lines for $Un/J_1=0$ and $1$ in Fig.~\ref{fig:phase}, 
where a small band gap closes and opens again as we change $Un/J_1$ between these values. 
If the system is located deep inside a region with nontrivial topology, 
a much larger band gap can be created, and edge modes can then be distinguished more clearly from bulk modes.  
}   
These results are consistent with the formation of chiral edge modes propagating in the $-y$ $(+y)$ direction 
at the left (right) boundary as expected from the bulk topological number $C_+=+1$. 

\begin{figure}
\begin{center}\includegraphics[width=0.9\textwidth]{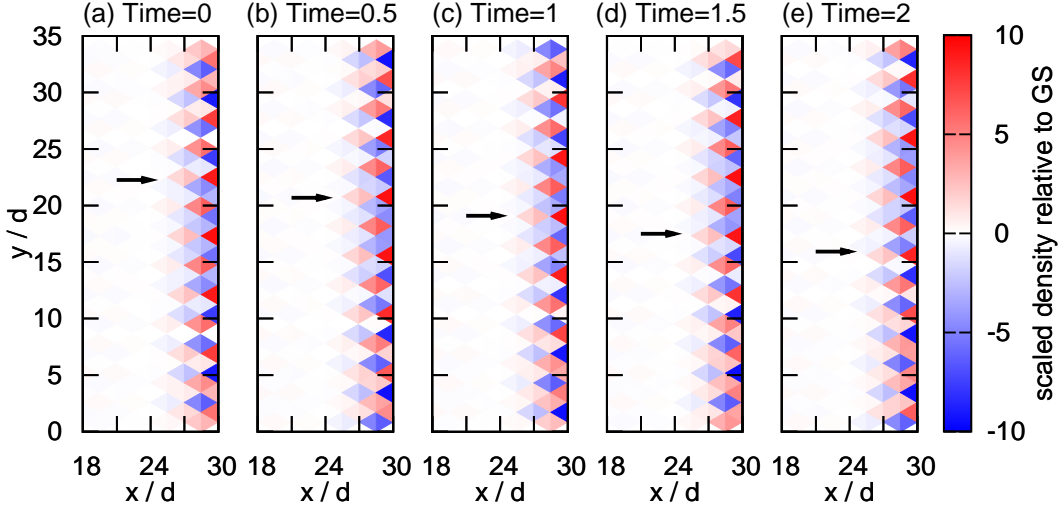}
\cutspace
\caption{\label{fig:if_strip}
Interference patterns of an edge matter wave with the background condensate 
at different times $J_1 t/\hbar=0,0.5,1,1.5,2$. 
The color shows the scaled density relative to that of the ground state, $\frac{U}{J_1\alpha} \left[ |\psi(x,y,t)|^2 -n_y|f_x|^2 \right]$, 
calculated from Eq.~\eqref{eq:density_interfere}. 
The center of each triangular pixel corresponds to a site of the honeycomb lattice. 
The edge matter wave is created by transferring a portion of the condensate 
into the edge mode with the momentum $k_y (\sqrt{3}d)/(2\pi)=-0.35$ 
(equivalent to 0.65 due to the periodicity of the Brillouin zone)
and the energy $E_{40}(k_y)/J_1=4.03$ as can be seen from Fig.~\ref{fig:spec_strip}(b). 
During the time evolution (a)-(e), the pattern propagates in the (negative) $y$ direction along the right edge 
with the phase velocity $E_{40}(k_y)/(k_y J_1)=-1.83\times \sqrt{3}d$; see, e.g., the propagation of an antinode with a positive variation (red) as indicated by the arrows. 
}\end{center}
\end{figure}


To detect the chiral edge modes between the two excitation bands, 
a high-frequency probe is required. 
This contrasts with the case of fermionic topological insulators, 
where edge modes cross the Fermi level and can be excited with infinitesimal energies. 
In ultracold-atom experiments, stimulated Raman transitions 
can be used to create excitations with desired momentum and frequency. 
Since edge modes are isolated from bulk modes in momentum and frequency in Fig.~\ref{fig:spec_strip}, 
Raman transitions can transfer a portion of a condensate selectively 
into a particular edge mode with momentum $k_y$ and frequency $\omega=E_j(k_y)/\hbar$, 
realizing an ``edge matter wave.''\footnote{
While the edge matter wave has an infinite lifetime within the BdG theory, 
it can acquire a finite lifetime due to collisions between quasiparticles and condensed particles (known as Beliaev damping \cite{Beliaev58}). 
The estimation of this lifetime however requires a detailed analysis of the collision processes, 
which is beyond the scope of the present paper. 
} 
The resulting condensate wave function is a coherent superposition 
of the background condensate and the edge matter wave, which is given by
\begin{equation}\label{eq:psi_interfere}
\psi(x,y,t)\approx \sqrt{n_y} e^{-i\mu t/\hbar} \left[  
 f_x + \alpha u_{xj}(k_y)e^{i(k_y y-\omega t)} + \alpha^* v_{xj}^*(k_y) e^{-i(k_y y-\omega t)} \right],
\end{equation}
where $\alpha$ is the complex amplitude of the edge mode. 
We assume $|\alpha|^2\ll 1$ to ensure that the linearization done in Eq.~\eqref{eq:GP_linear} to derive the BdG theory works. 
We relate the amplitude $\alpha$ to the microscopic process later. 
Since the edge mode has its weight mainly around an edge, 
we can expect that a density wave is formed along the edge as a result of the interference with the background condensate. 
Using Eq.~\eqref{eq:psi_interfere}, the scaled density profile relative to the ground state is calculated as
\begin{equation}\label{eq:density_interfere}
 \frac{U}{J_1} \left[ |\psi(x,y,t)|^2 -n_y|f_x|^2 \right] \approx \frac{Un_y}{J_1} \left[ \alpha z_{xj}(k_y) e^{i(k_yy-\omega t)}  + \mathrm{c.c.} \right]
\end{equation}
with $z_{xj}(k_y):=f_x^*u_{xj}(k_y) + f_x v_{xj}(k_y)$. 
We shift the origin of $t$ such that $\alpha$ becomes real. 
We plot Eq.~\eqref{eq:density_interfere} for different times in Fig.~\ref{fig:if_strip}. 
A density wave is indeed formed along the right edge, and it propagates in the negative $y$ direction 
with the phase velocity $E_{40}(k_y)/(k_y J_1)=-1.83\times \sqrt{3}d$ of the edge mode. 
We expect that such a propagating density wave can be used as a macroscopically enhanced experimental signature of an edge mode. 


In the above argument, we have kept the amplitude $\alpha$ of the edge matter wave undetermined. 
Here we determine $\alpha$ based on the microscopic process. 
This would help design an experimental setup for creating and observing an edge matter wave. 
A pair of Raman lasers with wave vectors $\kv_{1,2}$ and frequencies $\omega_{1,2}$ 
are prepared in such a manner that $\kv=\kv_1-\kv_2$ points in the $y$ direction [i.e., $\kv=(0,k_y,0)$], 
and that $\hbar\omega=\hbar(\omega_1-\omega_2)$ is resonant with the excitation energy $E_j(k_y)$ of the edge mode.  
These lasers induce the following time-dependent perturbation to the Hamiltonian:
\begin{equation}
 H'(t)=\sum_{(x,y)} \hbar \Omega \cos(k_y y-\omega t) a^\dagger (x,y) a (x,y),
\end{equation}
where the sum runs over positions $(x,y)$ of lattice sites, and $\Omega$ describes the strength of the atom-light coupling. 
This perturbation adds the following term to the RHS of the linearized GP equation \eqref{eq:GP_linear}: 
\begin{equation}
 \hbar\Omega \cos(k_y y-\omega t) \psi(x,y,t) 
 \approx \hbar\Omega \cos(k_y y-\omega t) \sqrt{n_y} f_x e^{-i\mu t/\hbar}.
\end{equation}
This describes a transfer of the condensate particles to the target edge mode; 
higher-order processes in which particles are further transferred to higher-energy states are neglected. 
In the presence of this term, we assume a solution of the form
\begin{equation}
 \phi(x,y,t)=\sqrt{n_y} e^{-i\mu t/\hbar} \left[ \alpha(t) u_{xj}(k_y) e^{i(k_y y-\omega t)} + \alpha^*(t) v_{xj}^*(k_y) e^{-i(k_y y-\omega t)}\right]. 
\end{equation}
The amplitude $\alpha(t)$ of the edge matter wave is found to obey 
\begin{equation}\label{eq:alpha_t}
 \left[ i\hbar \partial_t \alpha(t) \right] \tau_3 \wv_j(k_y)=\frac{\hbar\Omega}{2} \fv,
\end{equation}
where $\wv_j(k_y)=(\{u_{xj}(k_y)\},\{v_{xj}(k_y)\})^T$ as defined in Sec.~\ref{sec:strip} and $\fv:=(\{f_x\},\{f_x^*\})^T$. 
If Raman lasers are illuminated in the time interval $[0,\delta t]$, 
the integrated amplitude of the edge matter wave is obtained as
\begin{equation}
 \alpha(\delta t)=-i\frac{\Omega\delta t}{2} \wv_j^\dagger(k_y) \fv,
\end{equation}
where we have used $\wv_i^\dagger(k_y)\tau_3 \wv_j(k_y)=\delta_{ij}$ in solving Eq.~\eqref{eq:alpha_t}.
Since the GP ground state $\fv$ is extended over the strip and the edge-mode wave function $\wv_j(k_y)$ is localized only over a few sites around the edge, 
their overlap scales as $\wv_j^\dagger(k_y) \fv\sim 1/\sqrt{N_x}$. 
Using this result, we can tune $\Omega$ or $\delta t$ to achieve a desired amplitude $\alpha$. 

\subsection{Harmonic trap}


We next consider the case of a harmonic trap $V(x)=\frac{\kappa}{2} x^2$. 
The extended TF result \eqref{eq:Unfx2_TF} suggests that the condensate extends over the range $|x|<\RTF:=\sqrt{2U\nTF/\kappa}$. 
Integrating Eq.~\eqref{eq:Unfx2_TF} over this range, we find
\begin{equation}
 Un_y \approx \int_{-\RTF}^{\RTF} \frac4{3d} dx (\RTF^2-x^2) = \frac{8\kappa \RTF^3}{9d}=\frac{16\RTF}{9d} U\nTF.  
\end{equation}
To achieve the desired values of $U\nTF$ and $\RTF$, we can thus set the input parameters as
\begin{equation}\label{eq:kappa_Uny}
 \frac{\kappa}{2} = \frac{U\nTF}{\RTF^2}, ~~
 Un_y = \frac{16  \RTF}{9d} U\nTF.
\end{equation}

\begin{figure}
\begin{center}\includegraphics[width=0.9\textwidth]{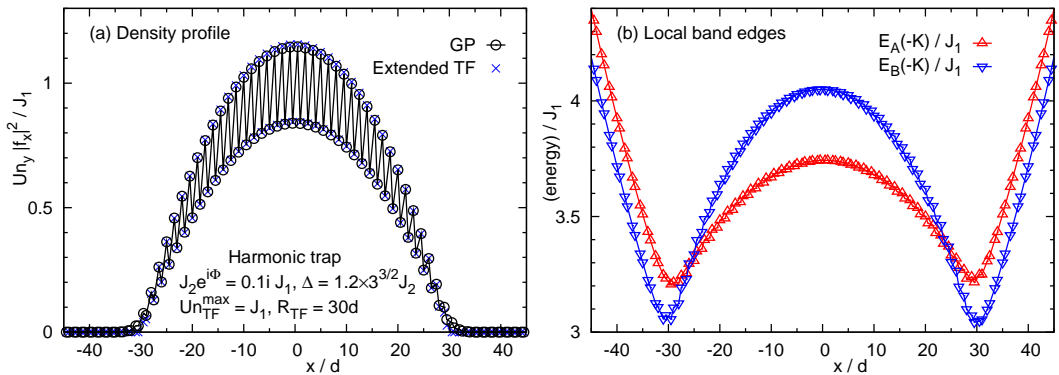}
\cutspace
\caption{\label{fig:gs_trap}
(a) Density profile (scaled by $U/J_1$) of the ground state in a harmonic trap. 
The data for the GP ground state are compared with the extended TF result \eqref{eq:Unfx2_TF}. 
The model parameters are chosen so that the case of Fig.~\ref{fig:bands}(c) 
is realized around the center of the trap. 
Calculations are performed in a sufficiently wide strip (with $N_x=160$) 
so that the effect of the strip edges is negligible. 
(b) Local band edges $E_{A/B}(-\Kv)$ calculated from the density profile in (a). 
For each $x\in X$, $E_X(-\Kv)$ is calculated 
by substituting the local density $Un_y|f_x|^2$ and the local potential $V(x)-\mu$ 
into $2Unf_X^2$ and $-\mu$, respectively, in Eq.~\eqref{eq:EXpmK};  
$E_{\bar{X}}(-\Kv)$ is calculated with a similar procedure using the average density at the two neighboring sites. 
}\end{center}
\end{figure}


We set $(J_2e^{i\Phi}/J_1,\Delta/J_2,U\nTF/J_1)=(0.1i,1.2\times 3^{3/2},1)$  
so that the case of Fig.~\ref{fig:bands}(c) (with nontrivial topology $C_+=+1$) 
is realized around the center of the trap. 
The scaled density profile $Un_y |f_x|^2 / J_1$ of the GP ground state in Fig.~\ref{fig:gs_trap}(a) 
indeed shows the maximum of near unity at the center; 
the oscillating pattern of the profile agrees well with the extended TF result \eqref{eq:Unfx2_TF}. 
As we move away from the center, the density decreases towards zero, 
and the case of Fig.~\ref{fig:bands}(a) (with trivial topology $C_+=0$) is expected to be realized for $|x|>\RTF$. 
This indicates that topological boundaries appear inside the condensate, 
assuming local homogeneity as in the semiclassical approach. 
To locate such boundaries, we plot in Fig.~\ref{fig:gs_trap}(b) the local band edges $E_X(-\Kv)$ with $X=A,B$, 
which are calculated by substituting the density profile of Fig.~\ref{fig:gs_trap}(a) into Eq.~\eqref{eq:EXpmK}. 
The two energies $E_{A/B}(-\Kv)$ indeed crosses at $x/d\approx -26$ and $24.5$, 
where topological boundaries are expected to be formed. 

\begin{figure}
\begin{center}\includegraphics[width=0.9\textwidth]{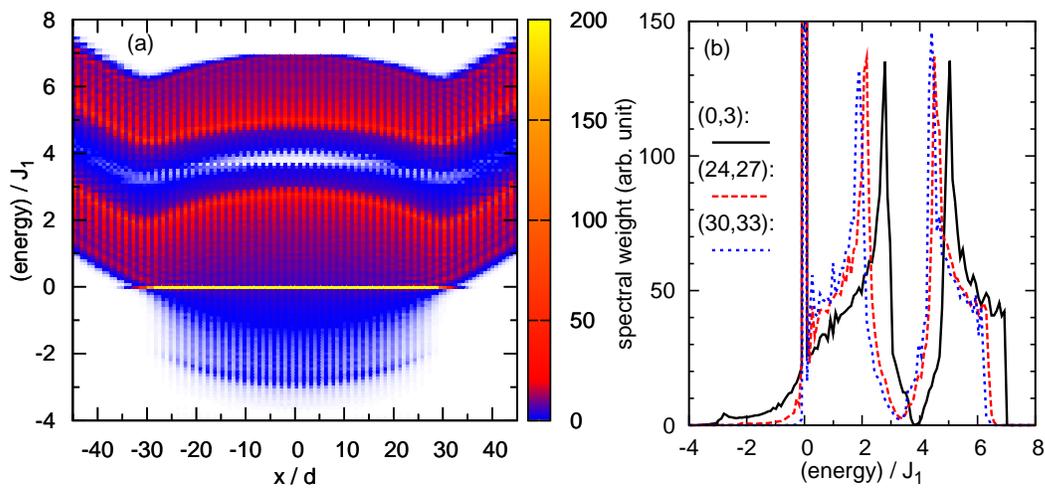}
\cutspace
\caption{\label{fig:spec_trap}
(a) Local spectral weight $\rho(x,x,\omega)=\sum_{k_y}\rho(x,x,k_y,\omega)$. 
A BdG calculation is performed using the GP ground state displayed in Fig.~\ref{fig:gs_trap}, 
and the spectral weight $\rho(x,x,k_y,\omega)$ is calculated with Eq.~\eqref{eq:rho_xxky}. 
(b) Integrated local spectral weight $\sum_{x\in I} \rho(x,x,\omega)$ for small intervals $I=(0,3d)$, $(24d,27d)$, and $(30d,33d)$. 
}\end{center}
\end{figure}


To see whether edge modes of the topological origin appear at such boundaries, 
we plot in Fig.~\ref{fig:spec_trap}(a) the local spectral weight $\rho(x,x,\omega)=\sum_{k_y}\rho(x,x,k_y,\omega)$. 
The formation of an excitation gap with the vanishing spectral weight is seen around the center $x=0$. 
As we move away from the center, the gap gradually closes as expected from the semiclassical approach. 
However, the reopening of the gap as in Fig.~\ref{fig:gs_trap}(b) is not seen in this figure. 
This is more clearly seen in the integrated spectral weight for small intervals shown in Fig.~\ref{fig:spec_trap}(b). 
For $I=(0,3d)$, the formation of a gap with the vanishing spectral weight can be seen. 
For $I=(24d,27d)$ and $(30d,33d)$, by contrast, the spectral weight has a tiny but non-vanishing value at the valley, 
indicating a gapless nature. 
This indicates the breakdown of the semiclassical approach. 
A possible reason for it is as follows. 
The local band edges calculated with the semiclassical approach in Fig.~\ref{fig:gs_trap}(b) indicate that 
not only the size of the gap but also its location (in energy) changes as a function of the position $x$. 
Therefore, beyond the semiclassical picture, the energy gap for a particular position $x$ is easily penetrated 
by the states in the same energy range in the surrounding region. 
The observation of edge modes in a harmonic trap thus remains a challenging issue. 


\section{Summary and outlook}\label{sec:summary}


We have studied the topological properties of Bogoliubov excitation bands in BECs in optical lattices 
on the basis of a Bose-Hubbard extension of the Haldane model. 
We have shown that the topological properties of the Bloch bands in the noninteracting case are smoothly carried over 
to those of Bogoliubov excitation bands in the interacting case, 
and that the parameter ranges showing nontrivial topology enlarges with increasing the Hubbard interaction or the particle density. 
In the presence of sharp boundaries, chiral edge modes appear in the gap between topologically nontrivial excitation bands. 
We propose that Raman transitions can be used to coherently transfer a portion of the condensate into an edge mode, 
and that a density wave is formed along the edge due to an interference with the background condensate. 
This can be used as a macroscopically enhanced experimental signature of the edge mode. 
By contrast, our results for a harmonic trap show that edge states are substantially obscured and difficult to observe, 
as opposed to what is expected from a semiclassical picture. 


We expect that BECs in optical lattices offer a unique playground in the studies of band topology. 
The macroscopic nature of BECs can enhance signatures of a topological edge mode. 
The high controllability of BECs offers various methods of exciting particles to such edge modes. 
While we have considered Raman transitions in this paper, 
a trap quench can provide another useful method. 
While both the bulk and edge modes are excited by such a quench, 
the edge excitations may exhibit a distinct time evolution because of their chiral nature 
(see Ref.~\cite{Goldman13PNAS} for related numerical demonstrations for fermions). 
It will also be interesting to exploit the high controllability of optical lattices 
to design bosonic systems with different symmetries or dimensionality,  
where different topological classes can be explored as expected from the studies of fermions \cite{Schnyder08,Kitaev09}. 


{\it Note added.}---Recently, we became aware of two independent works \cite{Engelhardt15,Bardyn15}, 
where nontrivial topology of Bogoliubov excitation bands and associated edge states were discussed in different bosonic systems. 
Engelhardt and Brandes \cite{Engelhardt15} have considered a one-dimensional system with inversion symmetry, 
while Bardyn {\it et al.}\ \cite{Bardyn15} have considered a kagome vortex lattice 
with potential realization in nonlinear optical systems or exciton-polariton condensates. 

\bigskip 

The authors thank Yusuke Horinouchi and Ryuichi Shindou for useful discussions. 
S.\ F.\ acknowledges the enlightening lecture by Shuichi Murakami on {\it Berry Phase Physics and Topological Insulators} 
(Dept.\ of Physics, Univ.\ of Tokyo, 2013), from which this work was partly motivated. 
This work was supported by 
KAKENHI Grant Nos.\ 25800225 and 26287088 from the Japan Society for the Promotion of Science, 
a Grant-in-Aid for Scientific Research on Innovative Areas ``Topological Materials Science'' (KAKENHI Grant No.\ 15H05855), 
the Photon Frontier Network Program from MEXT of Japan, 
and the Mitsubishi Foundation. 


\section*{References}
\newcommand{\etall}{{\it et al.}}
\renewcommand{\PRL}[3]{Phys. Rev. Lett. {\bf #1}, \href{http://link.aps.org/abstract/PRL/v#1/e#2}{#2} (#3)}
\newcommand{\PRLp}[3]{Phys. Rev. Lett. {\bf #1}, \href{http://link.aps.org/abstract/PRL/v#1/p#2}{#2} (#3)}
\newcommand{\PRA}[3]{Phys. Rev. A {\bf #1}, \href{http://link.aps.org/abstract/PRA/v#1/e#2}{#2} (#3)}
\newcommand{\PRAp}[3]{Phys. Rev. A {\bf #1}, \href{http://link.aps.org/abstract/PRA/v#1/p#2}{#2} (#3)}
\newcommand{\PRAR}[3]{Phys. Rev. A {\bf #1}, \href{http://link.aps.org/abstract/PRA/v#1/e#2}{#2} (R) (#3)}
\newcommand{\PRB}[3]{Phys. Rev. B {\bf #1}, \href{http://link.aps.org/abstract/PRB/v#1/e#2}{#2} (#3)}
\newcommand{\PRBp}[3]{Phys. Rev. B {\bf #1}, \href{http://link.aps.org/abstract/PRB/v#1/p#2}{#2} (#3)}
\newcommand{\PRBR}[3]{Phys. Rev. B {\bf #1}, \href{http://link.aps.org/abstract/PRB/v#1/e#2}{#2} (R) (#3)}
\newcommand{\PRE}[3]{Phys. Rev. E {\bf #1}, \href{http://link.aps.org/abstract/PRE/v#1/e#2}{#2} (#3)}
\newcommand{\PRX}[3]{Phys. Rev. X {\bf #1}, \href{http://dx.doi.org/10.1103/PhysRevX.#1.#2}{#2} (#3)}
\renewcommand{\RMP}[3]{Rev. Mod. Phys. {\bf #1}, \href{http://link.aps.org/abstract/RMP/v#1/e#2}{#2} (#3)}
\newcommand{\arXiv}[1]{arXiv:\href{http://arxiv.org/abs/#1}{#1}}
\newcommand{\condmat}[1]{cond-mat/\href{http://arxiv.org/abs/cond-mat/#1}{#1}}
\renewcommand{\JPSJ}[3]{J. Phys. Soc. Jpn. {\bf #1}, \href{http://journals.jps.jp/doi/abs/10.1143/JPSJ.#1.#2}{#2}(#3)}
\newcommand{\PTPS}[3]{Prog. Theor. Phys. Suppl. {\bf #1}, \href{http://ptp.ipap.jp/link?PTPS/#1/#2/}{#2} (#3)}
\newcommand{\hreflink}[1]{\href{#1}{#1}}


\begin{thebibliography}{10}

\bibitem{Hasan10}
 M. Z. Hasan and C. L. Kane, \RMP{82}{3045}{2010}. 
\bibitem{Qi11}
 X.-L. Qi and S.-C. Zhang, \RMP{83}{1057}{2011}. 
\bibitem{Bernevig13}
 B. A. Bernevig and T. L. Hughes, {\it Topological Insulators and Topological Superconductors} 
 (Princeton University Press, Princeton, NJ. 2013). 

\bibitem{vonKlitzing86}
 K. von Klitzing, \RMP{58}{519}{1986}. 
\bibitem{Thouless82}
 D. J. Thouless, M. Kohmoto, M. P. Nightingale, and M. den Nijs, \PRL{49}{405}{1982}; 
 M. Kohmoto, Ann. Phys. {\bf 160}, 343 (1985). 
 
\bibitem{Halperin82}
 B. I. Halperin, \PRBp{25}{2185}{1982}. 
\bibitem{Hatsugai93}
 Y. Hatsugai, Phys. Rev. Lett. 71, 3697 (1993).

\bibitem{Kane05}
 C. L. Kane and E. J. Mele, \PRL{95}{226801}{2005}. 
\bibitem{Bernevig06}
 B. A. Bernevig, T. L. Hughes, and S. C. Zhang, Science {\bf 314}, 1757 (2006).
\bibitem{Konig07}
 M. K\"onig, S. Wiedmann, C. Brune, A. Roth, H. Buhmann, L. W. Molenkamp, X. L. Qi, and S. C. Zhang, Science {\bf 318}, 766 (2007).

\bibitem{Fu07}
 L. Fu, C. L. Kane, and E. J. Mele, \PRL{98}{106803}{2007}. 
\bibitem{Moore07}
 J. E. Moore and L. Balents, \PRBR{75}{121306}{2007}. 
\bibitem{Roy09}
 R. Roy, \PRB{79}{195322}{2009}. 
\bibitem{Hsieh08}
 D. Hsieh, D. Qian, L. Wray, Y. Xia, Y. S. Hor, R. J. Cava, and M. Z. Hasan, Nature (London) {\bf 452}, 970 (2008).

\bibitem{Read00}
 N. Read and D. Green, \PRBp{61}{10267}{2000}. 
\bibitem{Ivanov01}
 D. A. Ivanov, \PRLp{86}{268}{2001}. 

\bibitem{Schnyder08}
 A. P. Schnyder, S. Ryu, A. Furusaki, and A. W. W. Ludwig, \PRB{78}{195125}{2008}. 
\bibitem{Kitaev09}
 A. Kitaev, AIP Conf. Proc. {\bf 1134}, 22 (2009). 

\bibitem{Dalibard11}
 J. Dalibard, F. Gerbier, G. Juzeli$\bar{\mathrm{u}}$nas, and P. \"Ohberg, \RMP{83}{1523}{2011}. 
\bibitem{Goldman13}
 N. Goldman, G. Juzeli$\bar{\mathrm{u}}$nas, P. \"Ohberg, I. B. Spielman, Rep. Prog. Phys. {\bf 77}, 126401 (2014). 

\bibitem{Lin09}
 Y.-J. Lin, R. L. Compton, K. Jinm\'enez-Garc\'ia, J. V. Porto, and I. B. Spielman, Nature {\bf 462}, 628 (2009). 

\bibitem{Aidelsburger13} 
 M. Aidelsburger, M. Atala, M. Lohse, J. T. Barreiro, B. Paredes, and I. Bloch, \PRL{111}{185301}{2013}.   
\bibitem{Miyake13} 
 H. Miyake, G. A. Siviloglou, C. J. Kennedy, W. C. Burton, and W. Ketterle, \PRL{111}{185302}{2013}. 
\bibitem{Aidelsburger15}
 M. Aidelsburger, M. Lohse, C. Schweizer, M. Atala, J. T. Barreiro, 
 S. Nascimbene, N. R. Cooper, I. Bloch, and N. Goldman, Nat. Phys. {\bf 11}, 162 (2015). 
\bibitem{Kennedy15}
 C. J. Kennedy, W. C. Burton, W. C. Chung, and W. Ketterle, Nat. Phys. {\bf 11}, 859 (2015). 

\bibitem{Celi14}
 A. Celi, P. Massignan, J. Ruseckas, N. Goldman, I. B. Spielman, G. Juzeli$\bar{\text{u}}$nas, and M. Lewenstein, \PRL{112}{043001}{2014}. 
\bibitem{Mancini15}
 M. Mancini, G. Pagano, G. Cappellini, L. Livi, M. Rider, J. Catani, C. Sias, P. Zoller, M. Inguscio, M. Dalmonte, and L. Fallani, Science {\bf 349}, 1510 (2015). 
\bibitem{Stuhl15}
 B. K. Stuhl, H.-I Lu, L. M. Aycock, D. Genkina, and I. B. Spielman, Science {\bf 349}, 1514 (2015).
 

\bibitem{Haldane88} 
 F. D. M. Haldane, \PRLp{61}{2015}{1988}. 

\bibitem{Jotzu14} 
 G. Jotzu, M. Messer, R. Desbuquois, M. Lebrat, T. Uehlinger, D. Greif, and T. Esslinger, Nature {\bf 515}, 237 (2014).  
\bibitem{Shao08}
 L. B. Shao, S.-L. Zhu, L. Sheng, D. Y. Xing, and Z. D. Wang, \PRL{101}{246810}{2008}. 
\bibitem{Stanescu09}
 T. D. Stanescu, V. Galitski, J. Y. Vaishnav, C. W. Clark, and S. Das Sarma, \PRA{79}{053639}{2009}. 
\bibitem{Alba11}
 E. Alba, X. Fernandez-Gonzalvo, J. Mur-Petit, J. K. Pachos, and J. J. Garcia-Ripoll, \PRL{107}{235301}{2011}. 
\bibitem{Goldman13NJ}
 N. Goldman, E. Anisimovas, F. Gerbier, P. Ohberg, I. B. Spielman, and G. Juzeli$\bar{\mathrm{u}}$nas, 
 New J. Phys. {\bf 15}, 013025 (2013)
\bibitem{Zheng14}
 W. Zheng and H. Zhai, \PRAR{89}{061603}{2014}.
 
\bibitem{Duca15}
 L. Duca, T. Li, M. Reitter, I. Bloch, M. Schleier-Smith, and U. Schneider, Science {\bf 347}, 288 (2015)
 
\bibitem{Haldane08}
 F. D. M. Haldane and S. Raghu, \PRL{100}{013904}{2008}; 
 S. Raghu and F. D. M. Haldane, \PRA{78}{033834}{2008}. 

\bibitem{ZhangNiu10}
 C. Zhang and Q. Niu, \PRA{81}{053803}{2010}. 

\bibitem{Wang09}
 Z. Wang, Y. Chong, J. D. Joannopoulos, M. Solja\v{c}i\'c, Nature (London) {\bf 461}, 772 (2009).

\bibitem{Hafezi13}
 M. Hafezi, S. Mittal, J. Fan, A. Migdall, and J. M. Taylor, Nature Photonics {\bf 7}, 1001 (2013). 

\bibitem{Rechtsman13}
 M. C. Rechtsman, J. M. Zeuner, Y. Plotnik, Y. Lumer, D. Podolsky, F. Dreisow, S. Nolte, M. Segev, and A. Szameit, 
 Nature {\bf 496}, 196 (2013). 
 
\bibitem{Prodan09}
 E. Prodan and C. Prodan, \PRL{103}{248101}{2009}. 
\bibitem{Berg11}
 N. Berg, K. Joel, M. Koolyk, and E. Prodan, \PRE{83}{021913}{2011}. 

\bibitem{Susstrunk15}
 R. Susstrunk and S. D. Huber, Science {\bf 349}, 47 (2015). 
 
\bibitem{Shindou13}
 R. Shindou, R. Matsumoto, S. Murakami, and J.-i. Ohe, \PRB{87}{174427}{2013};  
 R. Shindou, J.-i. Ohe, R. Matsumoto, S. Murakami, and E. Saitoh, \PRB{87}{174402}{2013}. 
\bibitem{Shindou14}
 R. Shindou and J.-i. Ohe, \PRB{89}{054412}{2014}. 
\bibitem{Romhanyi15}
 J. Romh\'anyi, K. Penc, and R. Ganesh, Nat. Comm. {\bf 6}, 6805 (2015). 

\bibitem{Karzig15}
 T. Karzig, C.-E. Bardyn, N. Lindler, G. Refael, \PRX{5}{031001}{2015}; 
 C.-E. Bardyn, T. Karzig, G. Refael, T. C. H. Liew, \PRBR{91}{161413}{2015}. 


\bibitem{Vasic15}
 I. Vasic, A. Petrescu, K. Le Hur, and W. Hofstetter, \PRB{91}{094502}{2015}. 

\bibitem{HeJing11}
 J. He, Y.-H. Zong, S.-P. Kou, Y. Liang, and S. Feng, \PRB{84}{035127}{2011}.
\bibitem{Prychynenko15}
 D. Prychynenko and S. D. Huber, \arXiv{1410.2001}.
\bibitem{ZhengWei15}
 W. Zheng, H. Shen, Z. Wang, and H. Zhai, \PRB{91}{161107}{2015}. 

\bibitem{Fouet01}
 J. B. Fouet, P. Sindzingre, and C. Lhuillier, Eur. Phys. J. B {\bf 20}, 241 (2001). 
\bibitem{Okuma10}
 S. Okuma, H. Kawamura, T. Okubo, and Y. Motome, \JPSJ{79}{114705}{2010}. 
  
\bibitem{Pethick08}
 C. Pethick and H. Smith, {\it Bose-Einstein Condensation in Dilute Bose Gases}, 2nd ed. (Cambridge University Press, New York, 2008).
\bibitem{Pitaevskii03}
 L. P. Pitaevskii and S. Stringari, {\it Bose-Einstein Condensation} (Clarendon Press, 2003). 
 
\bibitem{Colpa78}
 J. H. P. Colpa, Physica A 93, 327 (1978).
 
\bibitem{Fukui05}
 T. Fukui, Y. Hatsugai, and H. Suzuki, \JPSJ{74}{1674}{2005}. 

\bibitem{Ruprecht95}
 P. A. Ruprecht, M. J. Holland, K. Burnett, and M. Edwards, \PRA{51}{4704}{1995}. 

\bibitem{Buchhould12}
 M. Buchhould, D. Cocks, and W. Hofstetter, \PRA{85}{063614}{2012}. 

\bibitem{Gaunt13}
 A. L. Gaunt, T. F. Schmidutz, I. Gotlibovych, R. P. Smith, and Z. Hadzibabic, \PRL{110}{200406}{2013}. 

\bibitem{Beliaev58}
 S. T. Beliaev, Soviet Physics JETP {\bf 7}, 298 (1958); 299 (1958). 

\bibitem{Goldman13PNAS}
 N. Goldman, J. Dalibard, A. Dauphin, F. Gerbier, M. Lewenstein, P. Zoller, and I. B. Spielman, 
 PNAS {\bf 110}, 6736 (2013)

\bibitem{Engelhardt15}
 G. Engelhardt and T. Brandes, \PRA{91}{053621}{2015}. 
\bibitem{Bardyn15}
 C.-E. Bardyn, T. Karzig, G. Refael, and T. C. H. Liew, \arXiv{1503.08824}. 

\end{thebibliography}
\end{document}